\documentclass[final,2p,times,twocolumn]{elsarticle}
\usepackage{amssymb}
\usepackage[
  a4paper,
  left=1.5cm,
  right=1.5cm,
  top=2.0cm,
  bottom=2.0cm
]{geometry}
\usepackage{array}
\usepackage{booktabs}
\usepackage{tabularx}
\usepackage{amsmath}
\usepackage{natbib}
\affiliation[inst1]{organization={Department of Physics, University of Pretoria},%Department and Organization
            addressline={Lynnwood Road, Hatfield}, 
            city={Pretoria},
            postcode={0028}, 
            country={South Africa}}

\author[inst1]{Francois Conradie}
\author[inst1]{Bertus van Heerden\fnref{inst2}}
\fntext[inst2]{Present address: Department of Chemistry, University of Zululand, 1 Main Road, Vulindlela, kwaDlangezwa, 3886, South Africa}
\author[inst1]{Michal Gwizdala\fnref{inst3}}
\fntext[inst3]{Present address: The Institute of Photonic Sciences (ICFO), Barcelona, Spain}
\author[inst1]{Tjaart P. J. Krüger\corref{cor1}}
\cortext[cor1]{Corresponding author.}
\ead{tjaart.kruger@up.ac.za}

\journal{Biochimica et Biophisica Acta - Bioenergetics}

\begin{document}

\begin{frontmatter}

\title{Size-Dependent Fluorescence Kinetics Reveal Contributions of Intrinsic Quenching and Singlet–Triplet Annihilation during LHCII Aggregation}

\begin{abstract}
%% Text of abstract
Aggregation of the main antenna complex of higher plants, Light-Harvesting Complex II (LHCII), is widely used as an in vitro model for energy-dependent quenching (qE), yet fluorescence reduction in aggregates is frequently interpreted without a quantitative separation of intrinsic quenching from excitation-induced annihilation. Here, we address this ambiguity by directly correlating aggregate size, concentration, steady-state fluorescence intensity, and decay kinetics during controlled, incremental aggregation of isolated LHCII. By combining fluorescence correlation spectroscopy (FCS) with time-correlated single-photon counting (TCSPC) in a unified experimental framework, we monitored structural and photophysical changes in real time as detergent removal drives biphasic aggregation. We quantified the aggregate composition from the particle concentrations, enabling direct scaling of the absorption cross-section with aggregate size. The average fluorescence lifetime decreased semi-logarithmically with increases in hydrodynamic radius, whereas steady-state fluorescence intensities deviated strongly from this trend. Intensity-dependent measurements and steady-state kinetic modeling reveal that singlet-triplet annihilation (STA) emerges at moderate excitation intensities and rapidly becomes the dominant contributor to fluorescence quenching, even for relatively small aggregates. In contrast, intrinsic quenching increases more gradually with aggregate size. By quantitatively disentangling intrinsic excitation quenching from annihilation processes, this work demonstrates that STA can govern the apparent photophysical response of aggregated LHCII across excitation regimes commonly considered non-annihilating. The size-dependent mechanistic framework presented here provides a basis for distinguishing intrinsic quenching from annihilation effects in aggregation-based studies of photosynthetic antenna complexes.

\end{abstract}

%%Graphical abstract
\begin{graphicalabstract}
\includegraphics[scale=0.4]{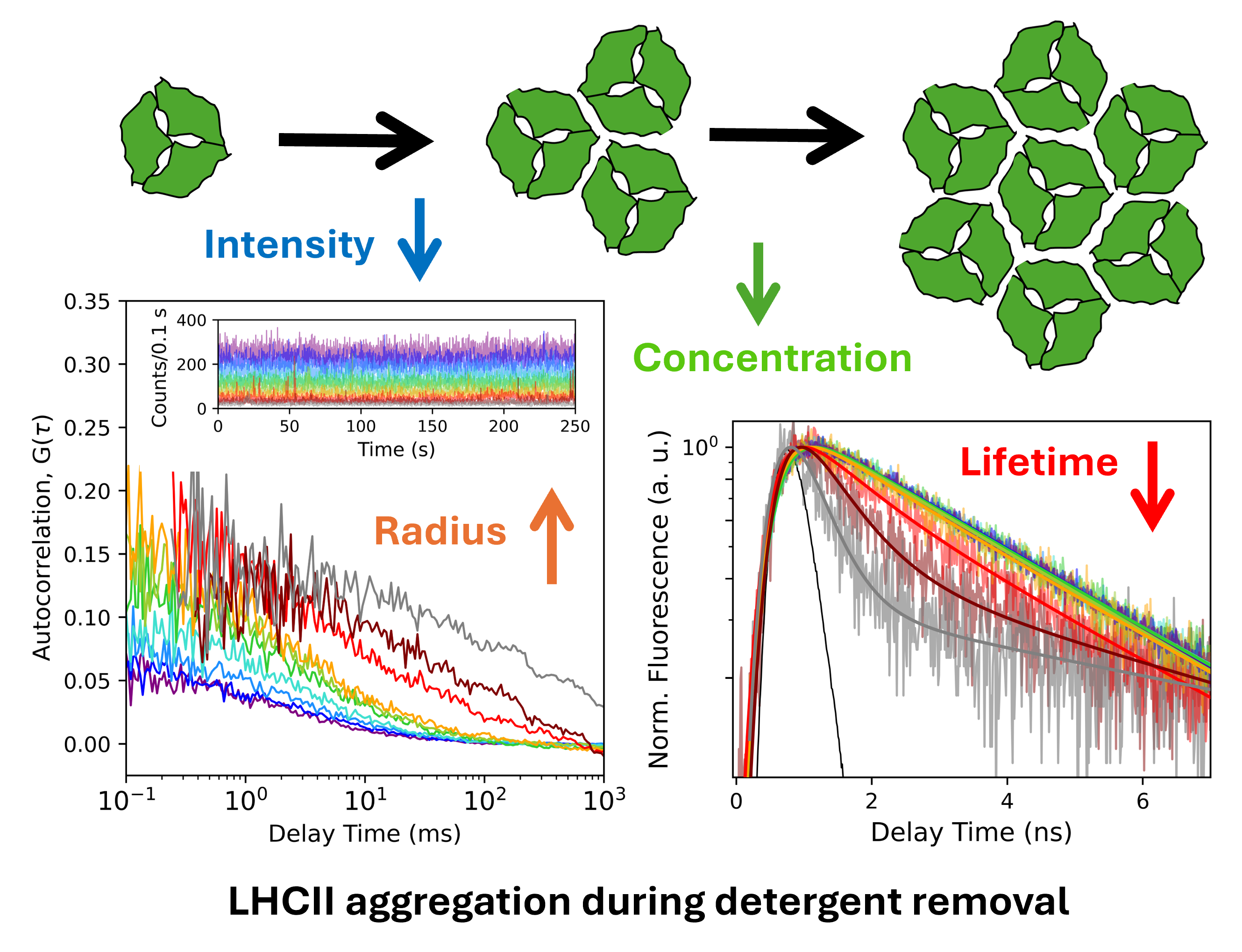}
\end{graphicalabstract}


\end{frontmatter}

%% \linenumbers

%% main text
\section{Introduction}
\label{sec:intro}
Life on Earth is sustained by the energy in solar radiation, which is captured and converted into usable chemical energy through photosynthesis. In green plants, photosynthesis begins in photosystems I and II (PSI and PSII), large, multi-subunit protein complexes embedded in the thylakoid membranes of chloroplasts. These complexes bind chlorophyll (Chl) and carotenoid (Car) pigments in well-defined positions and orientations, enabling absorption of most of the visible light in the solar spectrum and rapid transport of the excitation energy to the PSI and PSII reaction center complexes, where highly efficient charge separation takes place.
Proton accumulation in the thylakoid lumen establishes a pH gradient across the membrane, which, when sufficiently large, activates non-photochemical quenching (NPQ) \cite{Ruban2012}. \par
The main light-harvesting complex associated with PSII is LHCII, a heterotrimeric pigment--protein complex in which each subunit binds 14 Chls and 4 Cars \cite{Liu2004}. LHCII serves as the main antenna for light harvesting and is also central to rapid, reversible photoprotection in green plants \cite{Ruban2012}, known as qE, the energy-dependent component of NPQ. The interaction between excitonically coupled Chl clusters and Cars, particularly lutein, has frequently been explored as the cause of qE  \cite{Ruban2012, Ruban1994, Standfuss2005, Ruban2007, Johnson2009alteration, Chmeliov2016}, although Chl--Chl interactions have also been proposed \cite{Muller2010, Wahadoszamen2012}. Despite more than three decades of study, the molecular mechanisms that trigger qE remain unresolved. Current views generally ascribe the switching between light-harvesting and photoprotective states of LHCII to subtle conformational changes in individual complexes \cite{Ruban2007, 2vanOort2007, Kruger2012, Liguori2015}, modulated by the lumenal pH \cite{Ruban1992, Noctor1993, Ruban2012}, allosterically by the xanthophyll cycle \cite{Niyogi1998, Bose2008, Johnson2009Zea}, and the PsbS protein \cite{Li2000, Li2004, Wilson2024}. 

These molecular mechanisms involve the migration of delocalized singlet excitons to energy traps, characterized by the exciton diffusion length \cite{Bennet2018}, with trapping manifested as a pronounced shortening of the fluorescence lifetime.
Exciton trapping can also arise from the aggregation of LHCII complexes, providing an additional quenching pathway. Notably, the spectroscopic signals of \textit{in vitro} LHCII aggregates closely resemble those for LHCII under \textit{in vivo} qE conditions \cite{Ruban1992}. Consequently, LHCII aggregates are often used as model systems to study the molecular basis of qE. Moreover, some degree of LHCII oligomerization has been proposed to occur during NPQ, suggesting that LHCII aggregates may be a key component of qE \cite{Betterle2009, Johnson2011}.

LHCII aggregates can be readily prepared by isolating LHCII trimers in detergent micelles and incubating them with detergent-adsorbing resins (typically Bio-Beads SM-2) or diluting the complexes in a detergent-free buffer. Both methods decrease the detergent concentration well below its critical micelle concentration (CMC) to induce protein oligomerization. Detergent removal from a solution of isolated LHCII complexes with Bio-Beads for one to two hours is believed to lead to the complete aggregation of LHCII, necessary to obtain sufficiently large levels of quenching comparable to qE \cite{Muller2010, Lambrev2011}.\par

A major challenge in extracting kinetic information about quenchers formed in aggregated LHCII is that time-resolved techniques used for such studies can generate multiple excitons simultaneously within an aggregate, especially in large aggregates. Under these conditions, excitons themselves act as energy traps through exciton--exciton annihilation (EEA), a process that has been described in detail by theoretical models for molecular aggregates (see, e.g., \cite{Paillotin1979, Paillotin1983, vanGrondelle1985}). When two singlet excitons diffuse to neighboring Chls in LHCII, singlet-singlet annihilation (SSA) occurs, whereby one excitation is quenched. SSA in LHCII aggregates can be readily investigated using ultrafast transient-absorption spectroscopy \cite{Barzda2001} and has been used as a marker of aggregation, providing information on the number of interconnected chromophores \cite{Ilioaia2008, Gillbro1988}.

An additional important factor is the relatively high yield of intersystem crossing of LHCII-bound Chls ($\approx31\%$) \cite{Kramer1980}, which produces long-lived triplet excitons that can react with O$_2$ to produce toxic singlet oxygen and other reactive oxygen species (ROS). In LHCII, however, nearly all Chl triplet states are transferred to neighboring Cars within $\sim1$ ps \cite{Peterman1995}, efficiently suppressing ROS formation. When a singlet exciton diffuses to a Chl adjacent to a Car already in a triplet excited state, singlet--triplet annihilation (STA) occurs. In this process, the singlet exciton energy is transferred to the energetically lower Car triplet state, promoting it to a higher triplet level, after which it rapidly converts back to the lowest triplet state. The net effect is quenching of the singlet excitation. \par

Under pulsed excitation at high repetition rates ($\sim$100 kHz to MHz or higher) and sufficiently high intensities, triplet excited states are generally accumulated in light-harvesting complexes \cite{Gruber2015} and STA becomes a dominant excitation decay pathway \cite{Valkunas1991}. This arises from pulse separations being similar to or shorter than the decay time of Car triplet states in plant complexes. Under ambient conditions, the triplet lifetime of LHCII-bound Cars ranges from 2 to 4 $\mu$s \cite{Peterman1995, Schodel1999}, whereas in low-oxygen environments it increases to about 6 -- 9 $\mu$s \cite{Gruber2015}. Consequently, the accumulation of triplet states during pulsed excitation is governed by the ratio of the triplet lifetime to the time interval between successive excitation pulses. \par 

It is important to note that transient-absorption spectroscopy is typically performed at pulse repetition rates below $100$ kHz, under which conditions the accumulation of Car triplets in multichromophoric systems is generally negligible \cite{Valkunas2009}. In contrast, the possible contribution of STA in fluorescence-based measurements is often overlooked. One such widely used technique for studying NPQ in plants is time-correlated single-photon counting (TCSPC), which provides access to fluorescence lifetime information. Thermodynamic modeling of TCSPC fluorescence decay kinetics has shown that, when properly interpreted, these measurements can yield a surprisingly detailed picture of the quenching processes operative in LHCII aggregates  \cite{Gray2024}. However, TCSPC experiments typically employ high excitation pulse repetition rates, typically in the MHz range, and are therefore highly susceptible to STA.

It is particularly informative to correlate TCSPC with spatial or structural data to relate spectroscopic and dimensional information. In one such study, spatially-resolved TCSPC was combined with atomic force microscopy to investigate the influence of lipids on quenching dynamics in large, immobilized LHCII aggregates \cite{Adams2018}. Fluorescence correlation spectroscopy (FCS) is another dimensional technique that provides information about particle concentration, size, and interactions and has been used to determine the sizes of LHCII aggregates in solution \cite{Crepin2021}. However, FCS has not yet been combined with TCSPC to study size--kinetic relationships in LHCII aggregates.
\par

In this study, FCS and TCSPC were performed simultaneously to correlate the absolute fluorescence intensity and lifetime of LHCII complexes with their concentration and size during the transition from their trimeric light-harvesting state to oligomeric quenched states, monitoring the aggregate population in 5-minute intervals and quantifying the step-wise aggregation during gradual detergent removal. The experimental results were compared with simple kinetic and statistical models to quantify the amount of quenching, SSA, and STA during the time-dependent aggregation.

\section{Materials and Methods}
\label{sec:materials}
\subsection{Sample Preparation}
\label{subsec:sample_prep}
Thylakoid fragments were extracted from \textit{Spinacia oleracea} as described in Ref.~\cite{Caffiri2009}. LHCII trimers (OD$_{675}$ = 3.13) were extracted from the thylakoid fragments using sucrose gradient ultracentrifugation and isolated in 10 mM HEPES buffer at pH 7.5 with 0.03\% n-dodecyl-$\alpha$-D-maltoside ($\alpha$-DM) detergent as described in Ref.~\cite{Xu2015}, and absorption spectra were measured to confirm the purity of the sample (see Fig. S1). These stock samples were diluted 3000-fold to sub-nM concentrations for FCS measurements or 300-fold for bulk fluorescence decay measurements. Slow aggregation of the low concentration LHCII was induced by incubation of 200 mg/mL of Bio-Beads (SM-2 adsorbent resins, BioRad) using magnetic stirring at room temperature and in the dark (similar to Ref.~\cite{Lambrev2011}). Samples were extracted every 5 minutes for TCSPC--FCS measurements of LHCII after increasing increments of detergent removal. Data acquisition was started at least 10 minutes after extracting the first sample to allow for equilibration after each increment of detergent removal (see Fig. S2). This process was performed several times, with extra care taken to prevent photobleaching and to ensure consistent aggregation rates across repetitions. \par

\subsection{TCSPC-FCS}
\label{subsec:TCSPC-FCS}
FCS experiments were performed on a home-built single-molecule spectroscopy setup \cite{Kyeyune2019} modified for FCS. Light from a ps-pulsed supercontinuum laser (SuperK EVO, NKT Photonics, 20 MHz pulse repetition rate, $\sim$50 ps pulse duration) was filtered with a bandpass filter at 632.8 nm (Thorlabs, FL632.8-3). A dichroic beam splitter with a 649-nm cut-off wavelength (Semrock, BrightLine FF649-Di01-25x36) was used to direct the laser beam through a water-immersion objective (Nikon, 1.0 NA NIR Apo 60$\times$). The fluorescence photons from the samples were captured by the same objective and directed through a 75-$\mu$m confocal pinhole and a long-pass filter with a 650-nm cut-off (Thorlabs, FELH0650), before being focused onto a single-photon avalanche photodiode (SPAD) with an objective (Nikon, 0.1 NA NIR Apo 10$\times$). This detector, dubbed SPAD 1  (Excelitas SPCM-AQRH-16-TR), has a high photon detection efficiency of 75\% at 650 nm and an instrument response function (IRF) of 300--400 ps full-width half-maximum (FWHM), and was connected to a Becker \& Hickl\textsuperscript{\textregistered}, SPC-130EMN TCSPC module. It was used to record both the absolute fluorescence photon detection times for FCS and the detection microtimes for fluorescence decays.

To perform FCS calibration, the size of the confocal volume was measured by means of raster scans of $\sim$30 fluorescent microspheres (FluoSpheres, F8807, Thermo Fisher Scientific) with a piezoelectric nanopositioning stage (Nano-LP Series, Mad City Labs) at several positions along the beam's focus to obtain intensity distributions in the lateral and axial planes of the point spread function (see the Supplementary Methods and Figs. S3--S5). By fitting normal Gaussian functions to the intensity distributions, the lateral radius of the beam waist was found to be $\omega_0$ = 0.48 $\pm$ 0.04 $\mu$m and the axial radius $z_0$ = 3.1 $\pm$ 0.4 $\mu$m, giving an average confocal eccentricity of $k$ = 6 $\pm$ 1. To corroborate the measurement of the confocal volume, the diffusion time of a fluorescent dye (ATTO 647N free acid, Sigma-Aldrich) was measured using FCS to test whether the calibration parameters of the confocal volume were accurate, assuming a room-temperature diffusion coefficient of the dye of 41 $\mu$m$^2\cdot$s$^{-1}$. The excitation powers used for FCS averaged at 2.1 $\pm$ 0.2 $\mu$W, yielding an average beam waist intensity of 270 $\pm$30 W$\cdot$cm$^{-2}$ and average pulse energy of $\sim$0.1 pJ. The average laser pulse fluence at the beam waist was, therefore, 4.3 $\times$ $10^{13}$ photons$\cdot$pulse$^{-1}\cdot$cm$^{-2}$. This was a much lower intensity than what is commonly used for FCS of fluorescent dyes \cite{Enderlein2005, Schwille2000, Buschmann2007, Nicolau2009, Schwille2001, Gregor2005, Davis2006}, which was intended to investigate LHCII aggregation at what is considered to approximately be annihilation-free intensities for LHCII aggregates \cite{Barzda2001, Pascal2005, Johnson2009Zea, Ilioaia2008, Natali2016, Adams2018, Gray2024}. Furthermore, FCS was performed with the shortest possible measurement times, namely $\le$ 10 minutes, and the lowest viable excitation intensity, 270 $\pm$ 30 W$\cdot$cm$^{-2}$, to acquire autocorrelation curves exhibiting sufficiently low noise levels, but without causing too much delay before taking a measurement of the next extracted sample. The low excitation intensity and short measurement times also minimized the cumulative effect of photobleaching during the measurement. \par

To analyze time-resolved fluorescence decays, the IRF was measured by focusing the laser $\sim40$ $\mu$m above the glass--water interface of the microscope coverslip using a scattering medium (0.2\% fat-free milk). A deconvolution algorithm was used to fit the fluorescence decays with a biexponential model, using \textit{Full SMS} \cite{Botha2024}. Amplitude-averaged fluorescence lifetimes were calculated from the biexponential fits using Eq. \ref{eq:avg_lifetime}. The average intensities were calculated from the steady-state fluorescence time traces. \par

The contribution of slow photophysical processes, such as triplet dynamics, to the autocorrelation curves was minimized by considering correlation times between 100 $\mu$s and 1 s. Still, accurate fitting of the autocorrelation curves often required the inclusion of triplet blinking. Specifically, a combined triplet blinking and 3D diffusion model or a two-component 3D diffusion model was employed using PyCorrFit 1.1.7 \cite{PyCorrFit}. The best fits of the autocorrelation curves were obtained by applying the Levenberg-Marquardt algorithm to either a conventional one-component diffusion model (Eq. \ref{eq:FCS_model_T+3D}) or a two-component diffusion model (Eq. \ref{eq:FCS_model_3D+3D}). The amplitudes, $G(0)$, were used to extract the number of particles detected in the confocal volume during the course of a measurement, $N$, and the diffusion times, $\tau_D$, were converted to hydrodynamic radii, $R_H$, using Eq. \ref{eq:hydrodynamic_radius}. The hydrodynamic radius can be interpreted as the radius of a spherical particle with the same diffusion time as the measured macromolecule. The contribution of sub-millisecond photophysical processes did not significantly affect the values obtained for $N$ (see Supplementary Methods and Fig. S6). Background correction of $N$ was done by calculating $\chi^2$-factors (Eq. \ref{eq:Background_corr}). This ensured that the measurements of relative sample concentrations were accurate.

\subsection{Bulk TCSPC}
To corroborate the TCSPC--FCS measurements, standard TCSPC measurements of higher-concentration samples were performed using SPAD 2 (Micro Photon Devices, PD-050-CTE), which has an IRF of $38$ ps FWHM and a photon detection efficiency of 30\% at 670 nm. A lower and a higher excitation intensity were used for bulk time-resolved fluorescence measurements at the diffraction limit, namely 31 $\pm$ 1 W$\cdot$cm$^{-2}$ and 258 $\pm$ 6 W$\cdot$cm$^{-2}$ (corresponding to average pulse energies of 0.01 pJ and 0.1 pJ, respectively). \par
TCSPC measurements with SPAD 2 allowed for faster timing resolution, limited only by the laser pulse duration of $\sim50$ ps. Resolving $<100$-ps fluorescence decay components required fitting decays with three exponential components. Due to this detector's low photon detection efficiency, it was used only for bulk TCSPC measurements. \par

\section{Experimental Results}
\label{sec:results}
TCSPC--FCS enabled a detailed study of the aggregation of freely diffusing LHCII at room temperature, allowing us to simultaneously monitor the fluorescence count rate, \textbf{$F^{det}$}, amplitude-averaged fluorescence lifetime, \textbf{$\tau^{avg}$}, background-corrected average particle number, \textbf{$\langle N \rangle$}, and average hydrodynamic radius, \textbf{$R_H$}, of the complexes at increasing increments of aggregation.

\begin{figure}[hbt!]
    \centering
    \includegraphics[scale=0.70]{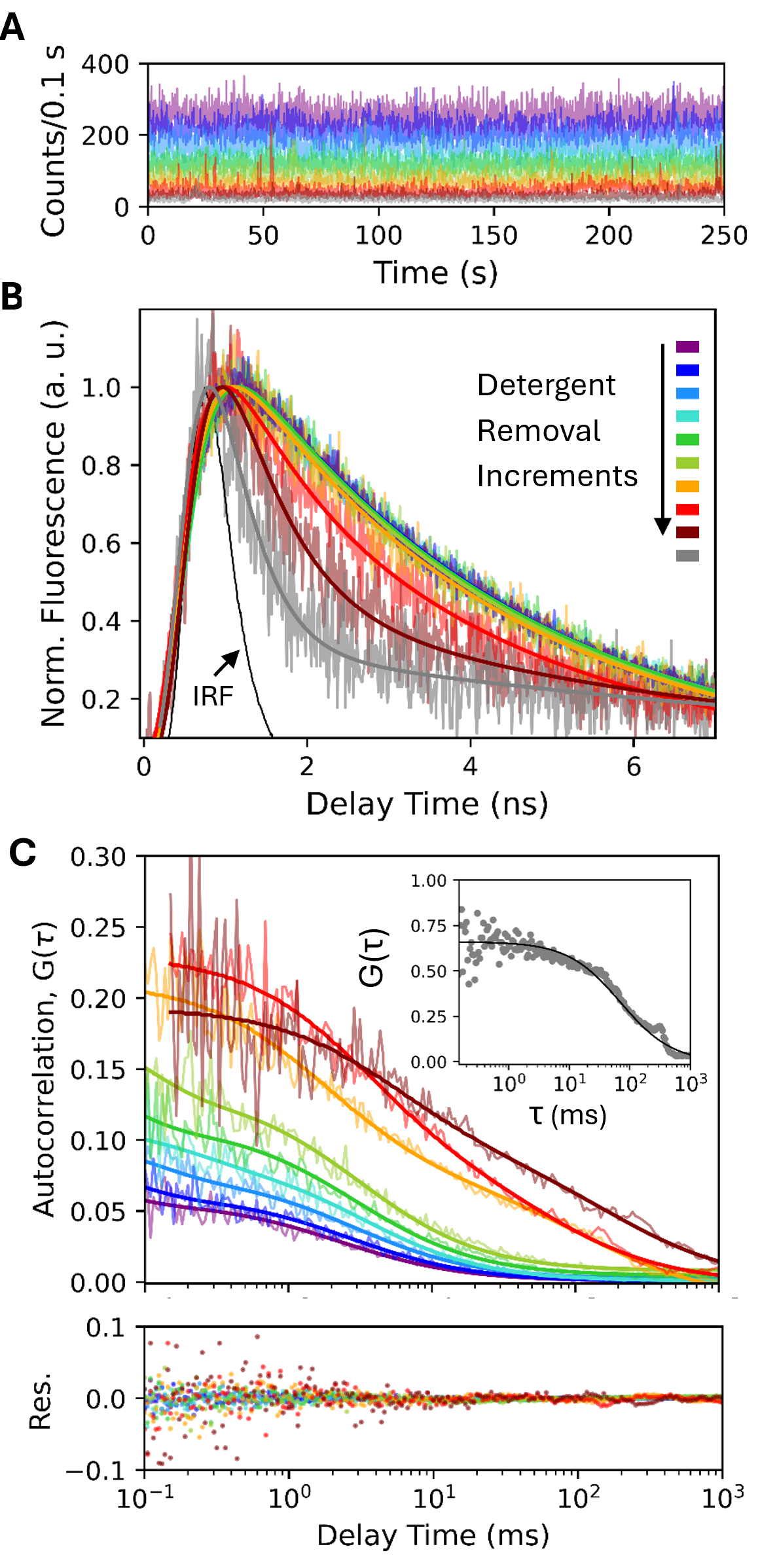}%width=\linewidth
    \caption{\textbf{A}. Fluorescence intensity traces of freely-diffusing LHCII recorded with SPAD 1 after increasing intervals of detergent removal. The colours indicate different samples extracted from the same solution incubated with Bio-Beads at 5-min intervals. \textbf{B}. Examples of fluorescence decays recorded in the same measurements using SPAD 1 (0.01 ns bin size). Biexponential fits are shown in black. \textbf{C}. Corresponding examples of raw autocorrelation curves of samples extracted, measured within 10 -- 50 minutes after extraction. $G(\tau)$-fits of the autocorrelation curves are shown in black; one or two 3D diffusion components were used. The inset shows an example of a $G(\tau)$-fit for very large LHCII aggregates. Fitting residuals are shown (bottom).}
    \label{fig:LHCII_decay_and_autocorrelation_dynamics}
\end{figure}

Example snippets of fluorescence intensity traces and corresponding ps--ns decay traces and autocorrelation curves at different stages of LHCII aggregation are shown in Figures \ref{fig:LHCII_decay_and_autocorrelation_dynamics}A, B, and C, respectively. Despite using a constant excitation intensity for all samples, the fluorescence count rates decreased by 80 -- 90\% over 40 -- 50 minutes of aggregation (Fig. \ref{fig:LHCII_decay_and_autocorrelation_dynamics}A) with concurrent decreases in the fluorescence lifetime (Fig. \ref{fig:LHCII_decay_and_autocorrelation_dynamics}B) as well as increases in diffusion times and decreases in detected particle numbers, as evidenced, respectively, by broader tails and higher $G(0)$ values in Figure \ref{fig:LHCII_decay_and_autocorrelation_dynamics}C.  \par

\begin{figure*}[hbt!]
    \centering
    \includegraphics[scale=0.44]{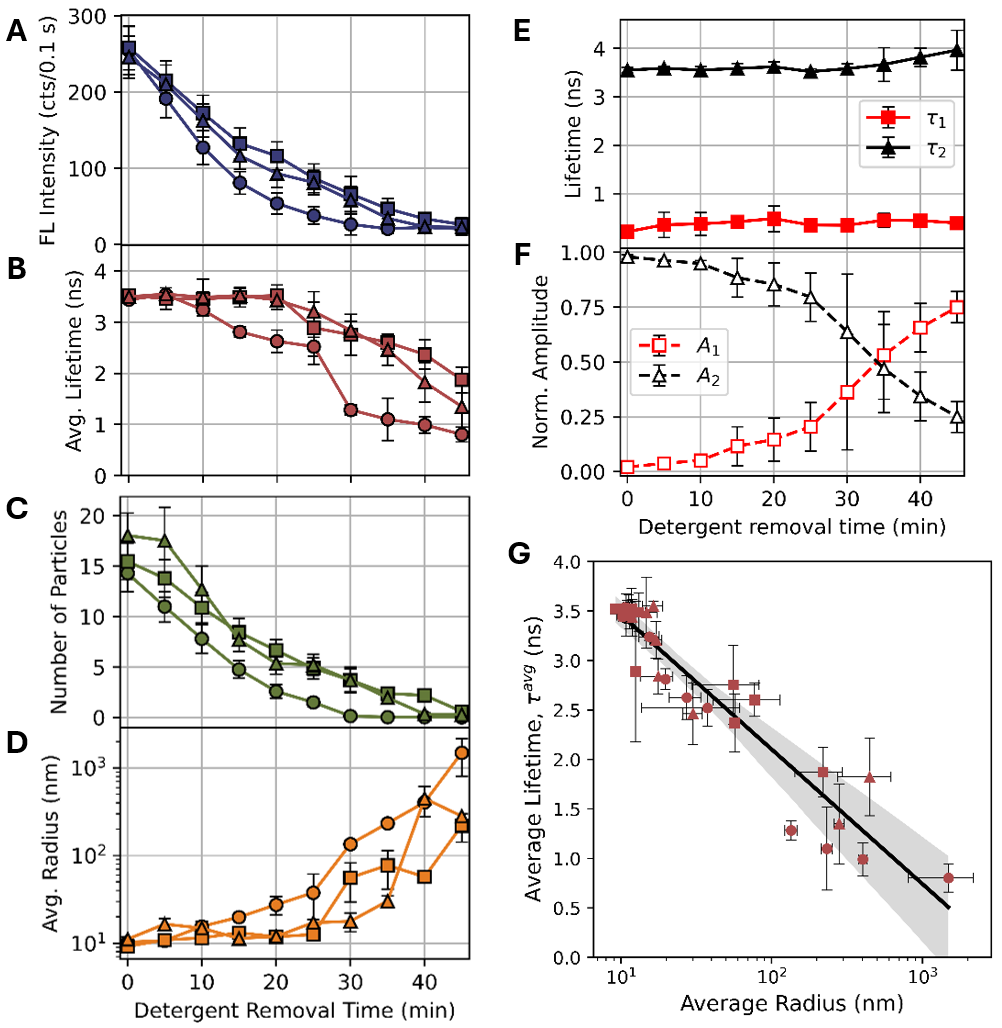}
    \caption{\textbf{A}. Average fluorescence intensities measured with SPAD 1 over detergent removal time of three repetitions of the incremental aggregation experiment, corresponding with examples shown in Fig. \ref{fig:LHCII_decay_and_autocorrelation_dynamics}A. \textbf{B}. Corresponding average fluorescence lifetimes from biexponential fits of decays corresponding with Fig. \ref{fig:LHCII_decay_and_autocorrelation_dynamics}B. \textbf{C}. Average background-corrected particle numbers corresponding with example autocorrelation curves shown in Fig. \ref{fig:LHCII_decay_and_autocorrelation_dynamics}C. \textbf{D}. Corresponding hydrodynamic radii obtained from autocorrelation diffusion times. Data points from three independent repetitions of the experiment can be distinguished by shape, viz., circles, squares, and triangles. \textbf{E}. Fluorescence lifetime components resolved by biexponential fitting. \textbf{F}. Normalized amplitudes of the two lifetime components in D. averaged over the three repetitions of the experiment, with $A_1$ corresponding with $\tau_1$ and $A_2$ with $\tau_2$.  \textbf{G}. Average fluorescence lifetimes compared with the average hydrodynamic radii. A semi-logarithmic linear regression (base $e$) with 95\% confidence interval is shown, with a gradient of $-0.59 \pm 0.04$ and $r^2 = 0.9$.}
    \label{fig:I_L_N_R_WT_LHCII}
\end{figure*}

These dynamics are quantified by the four simultaneously measured parameters \textbf{$F^{det}$}, \textbf{$\tau^{avg}$}, \textbf{$\langle N \rangle$}, and \textbf{$R_H$} shown in Figure \ref{fig:I_L_N_R_WT_LHCII}, all of which were averaged over consecutive 5-minute intervals. The fluorescence count rates (\textbf{$F^{det}$}) decreased most significantly during the first 30 minutes after Bio-Beads incubation (Fig. \ref{fig:I_L_N_R_WT_LHCII}A), while $\tau^{avg}$ remained initially constant and only started to shorten after 15 to 25 minutes elapsed (Fig. \ref{fig:I_L_N_R_WT_LHCII}B). Aggregation of LHCII was characterized by changes in both $\langle N \rangle$ (Fig. \ref{fig:I_L_N_R_WT_LHCII}C) and $R_H$ (Fig. \ref{fig:I_L_N_R_WT_LHCII}D). Independent repeats of the detergent removal process show statistically significant variations in the averaged fluorescence kinetics and aggregation quantities, as indicated by the three datasets in each of Figure \ref{fig:I_L_N_R_WT_LHCII}A--D. \par

Before detergent removal was initiated, the average hydrodynamic radius of LHCII trimers in $\alpha$-DM micelles was 10.2 $\pm$ 0.9 nm (based on an average diffusion time of 2.5 $\pm$ 0.2 ms and beam waist radius of 0.48 $\mu$m), which corresponds well to values reported in previous FCS studies \cite{Crepin2021, Iwai2013, Schaller2011}. Increased particle size heterogeneity was apparent after the onset of LHCII aggregation in the shape of the autocorrelation curves, which exhibited either one diffusion time with a large uncertainty or two diffusion times, as reflected by the extended long-delay tail end of the curves. Since the shorter diffusion times ($<$10 ms) correspond to the diffusion of trimers, the significantly longer diffusion times ($>$100 ms) are attributed to aggregates. Multiple diffusing molecular species were especially apparent for measurements performed after 25 -- 40 minutes of detergent removal, based on significant fitting uncertainties that improved when fitting the autocorrelation curves with a two-component diffusion model (see Fig. S7). The particle concentration, $\langle N \rangle$, already started to decrease within the first 5 minutes (Fig. \ref{fig:I_L_N_R_WT_LHCII}C), suggesting that a mixture of trimers and some aggregates started to form from the moment detergent removal started. During the first $\sim$20 minutes of detergent removal, $\langle N \rangle$ decreased significantly while $R_H$ increased only marginally (Fig. \ref{fig:I_L_N_R_WT_LHCII}D). This can be explained by the dominant trimer population during this period, giving rise to small average $R_H$ values, while the small initial value of $\langle N \rangle$ was more strongly affected than $R_H$ as the trimers clustered into small aggregates. 

Very large increases in $R_H$ were apparent after 25 minutes of the onset of detergent removal, with average radii of up to $\sim$1500 nm obtained after 45 minutes. However, extensive variations in $R_H$ along with only small variations in $\tau^{avg}$ over repetitions of 45 minutes of detergent removal showed that the preparation of sufficiently quenched LHCII aggregates can involve a large heterogeneity in the average aggregate size, corresponding to previous observations \cite{Adams2018, Crepin2021}. The relative uncertainties in $R_H$ and $\langle N \rangle$ increased from 8\% to 46\% from 0 to 45 minutes of detergent removal, again highlighting the extensive heterogeneity in aggregate sizes. Substantial variations in the aggregation process and aggregate-to-trimer ratios were also observed across repetitions, indicating the limited reproducibility of such time-dependent studies of LHCII aggregation. 
\par

For all fluorescence decays recorded with SPAD~1, two lifetime components could be resolved, averaging at $\tau_1 = 0.35$ $\pm$ $0.15$ ns and $\tau_2$ = $3.6$ $\pm$ $0.3$ ns, respectively, as shown in Figure \ref{fig:I_L_N_R_WT_LHCII}E. The short component is in the range of commonly reported values associated with energy-dependent quenching and aggregation of LHCII, while the long component corresponds to unquenched LHCII \cite{Johnson2009alteration, Chmeliov2016, Muller2010, Barzda2001, Gruber2015, vanOort2007, Miloslavina2008, Natali2016}. The limited photon budget precluded resolving a third lifetime component.  The significant shortening in $\tau^{avg}$ coincided with the large increases in $R_H$, resulting in a semi-logarithmic relationship between these two parameters (Fig. \ref{fig:I_L_N_R_WT_LHCII}G). \par

The amplitude of $\tau_1$ ($A_1$) only started to increase meaningfully after $\sim$10 minutes of detergent removal (Fig. \ref{fig:I_L_N_R_WT_LHCII}F), suggesting that the nonlinear relationship between $F^{det}$ and $\tau^{avg}$ during the first $\sim$20 minutes (Figs. \ref{fig:I_L_N_R_WT_LHCII}A and B) can only be partly explained by this component, which is ascribed to aggregation-related quenching. Therefore, another process is responsible for the decrease in the effective fluorescence yield, apparent in the steady decline of $F^{det}$, already within the first five minutes after Bio-Beads incubation. Resolving this requires a look into the different factors that affect the average fluorescence count rate of aggregating LHCII in solution, which is given by
\begin{equation}
    F^{det} = I_e \langle \sigma^{eff}\rangle \langle N \rangle \eta^{det} \langle \Phi^{eff} \rangle + \langle b \rangle,
    \label{eq:countrate}
\end{equation}

\noindent where $I_e$ is the average excitation intensity, $\sigma^{eff}$ is the average effective absorption cross-section, $\eta^{det}$ is the optical detection efficiency of the experimental setup, $\Phi^{eff}$ is the average effective fluorescence yield, and $b$ is the average background count rate, which was 190 $\pm$ 40 counts/s. The brackets ``$\langle \rangle$" denote time-averaging over the course of an FCS measurement. \par

The decreases in $F^{det}$ corresponded well with the decreases in $\langle N \rangle$, resulting in a strong linear correlation (see Fig. S8 A). This would not be expected if aggregation were the sole cause of the decrease in $\langle N \rangle$, since aggregation increases $\langle \sigma^{eff} \rangle$ while simultaneously decreasing $\langle N \rangle$ by the same factor at the ensemble level; under this assumption, the two effects should cancel out in Eq. \ref{eq:countrate}. Fig. S8 B shows that the count rates per molecule (CPM) calculated from $F^{det}$ and $\langle N \rangle$ stayed constant for a wide range of corresponding $\tau^{avg}$, except for highly quenched samples ($<$1 ns lifetimes). Hence, aggregation could not be characterized by CPM-values, but is apparent in the $\langle N \rangle$-values. The following aggregation condition can be imposed by these observations:
\begin{equation}
    \frac{\langle \sigma^{eff}\rangle}{\langle \sigma^{eff}_0 \rangle}\frac{\langle N \rangle}{\langle N_0 \rangle} = 1,
    \label{eq:aggregation_condition}
\end{equation}
where $\langle \sigma^{eff}_0 \rangle = \sigma$ is the average absorption cross-section of LHCII solubilized in 0.03\% $\alpha$-DM and $\langle N \rangle_0$ is the number of detected particles before detergent removal started. Given the average fluorescence count rate, $F=F^{det}-\langle b \rangle$, and the aggregation condition (Eq. \ref{eq:aggregation_condition}), the relative fluorescence count rate is equal to the relative effective fluorescence yield, i.e., 
\begin{equation}
    \frac{F}{F_0} = \frac{\langle \Phi^{eff} \rangle}{\langle \Phi^{eff}_0 \rangle},
\end{equation}
considering a constant $I_e$ and $\eta^{det}$.
Therefore, since the other parameters in Eq. \ref{eq:countrate} are constant, the immediate decreases in $F^{det}$ during detergent removal can only be explained by decreases in $\langle \Phi^{eff} \rangle$. \par

\begin{figure}[hbt!]
    \centering
    \includegraphics[scale=0.76]{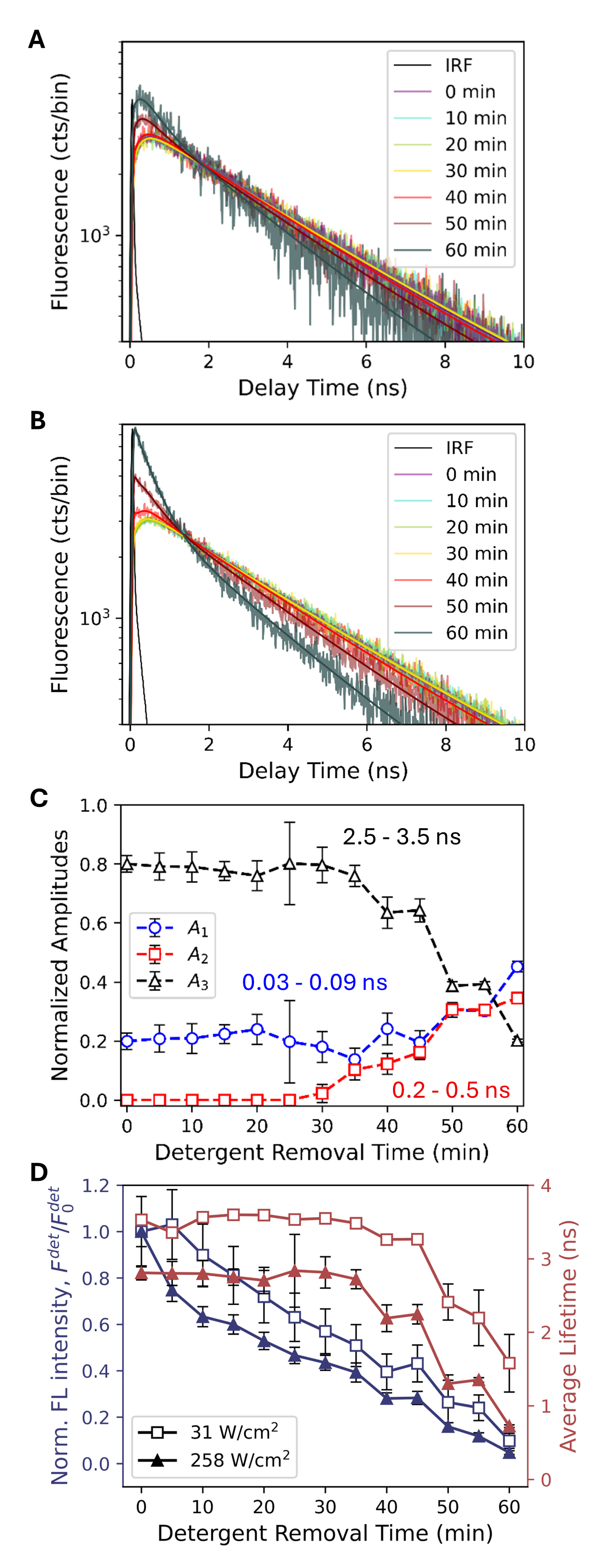}
    \caption{\textbf{A}. Fluorescence decays (0.01 ns bin sizes) of LHCII at 10-minute-long intervals of detergent removal time recorded with SPAD 2 ($\sim$38 ps IRF) at a lower excitation intensity (31 W$\cdot$cm$^{-2}$). \textbf{B}. Fluorescence decays recorded with SPAD 2 at a higher intensity (258 W$\cdot$cm$^{-2}$). \textbf{C}. Normalized amplitudes of the three resolved lifetime components at the higher excitation intensity, with corresponding average lifetime components indicated. \textbf{D}. Normalized average fluorescence intensities (left axis, blue) and average fluorescence lifetimes (right axis, red) of the same samples measured at 31 and 258 W$\cdot$cm$^{-2}$.}
    \label{fig:ps_L_STA_WT_LHCII}
\end{figure}

To investigate more closely the processes underlying the changes in $\langle \Phi^{eff} \rangle$, the TCSPC experiments were carried out at a sample concentration 10 times higher and using two different excitation intensities (31 and 258 W$\cdot$cm$^{-2}$ at the beam waist), the higher intensity being similar to that used for the FCS measurements. SPAD 2 was used for these measurements to resolve short-lifetime components accurately. The intensity-dependent study enabled the separation of aggregation-specific effects from intensity-dependent effects. Comparison of Figures \ref{fig:ps_L_STA_WT_LHCII}A and B shows that a higher excitation intensity gave rise to consistently faster decays and an enhanced initial fast decay, which is especially apparent for samples taken after long detergent removal times (i.e., larger aggregates). The higher-intensity data (258 W$\cdot$cm$^{-2}$ excitation, Fig. \ref{fig:ps_L_STA_WT_LHCII}B) could be sufficiently resolved with two lifetime components during the first 25 minutes of detergent removal, with values of 30 -- 90 ps and 2.5 -- 3.5 ns, respectively, while another lifetime component, 200 -- 500 ps, was apparent after 30 minutes. The corresponding normalized amplitudes are shown in Figure \ref{fig:ps_L_STA_WT_LHCII}C. Notably, the shortest lifetime component (30 -- 90 ps, with corresponding amplitude $A_1$ in Fig. \ref{fig:ps_L_STA_WT_LHCII}C, blue) was continually apparent at 258 W$\cdot$cm$^{-2}$ excitation, but it could not be resolved for the 31 W$\cdot$cm$^{-2}$ dataset, except after 50 minutes (see Fig. S9 A and C), indicating that it does not originate from aggregation-related quenching. Instead, this fast component corresponds to the EEA timescale, a process expected to exhibit significant dependence on excitation intensity and that occurs within tens of picoseconds \cite{Barzda2001, Gruber2015}.

The longest lifetime component, which corresponded at first with the fluorescence lifetime of unquenched LHCII complexes at 3.6 ns, had the greatest amplitude ($A_3$) for both excitation intensities except at the end of the detergent removal time. After 30 minutes, this component decreased in lifetime, down to 2.2 ns after 60 minutes for the higher intensity (see Fig. S9 B and D). The second lifetime ($\sim$0.4 ns) is associated with aggregation-induced quenching and had a smaller amplitude ($A_2$) than the fast component ($A_1$) for the higher-intensity data. Initially, $A_1\approx0.2$, meaning that 20\% of excitations were annihilated in LHCII trimers. After 60 minutes, $A_1\approx0.5$, being the greatest of the three amplitudes, meaning that 50\% of excitations were annihilated. It is clear that aggregation caused an increase in the EEA probability. For the sample exposed to a higher excitation intensity, EEA was the dominant quenching process, but EEA was negligible throughout most of the experiment at the lower excitation intensity, in which case only aggregation-related quenching could be resolved. \par

Normalized average fluorescence count rates are compared with amplitude-averaged fluorescence lifetimes in Figure \ref{fig:ps_L_STA_WT_LHCII}D, which exemplifies the deviation between declining LHCII fluorescence intensities and lifetimes during detergent removal. It is clear that $A_1$ did not increase significantly from the start of detergent removal, despite the marked decrease in absolute fluorescence count rates. This indicates that the time-resolved fluorescence kinetics was affected differently by annihilation than the steady-state fluorescence count rates, suggesting a nonlinear dependence of $A_1$ on the annihilation rate. This appeared to have been the case whether SPAD 1 or SPAD 2 was used, which means that EEA in LHCII trimers and aggregates was not fully resolved by either detector, likely because the excitation pulse duration ($\sim$50 ps) had been longer than the equilibration time of energy transfer processes in a trimer ($<$40 ps) \cite{Valkunas2009, Gruber2015}. \par
% Aggregation appears to have occurred a bit more slowly for these samples prepared at a higher LHCII concentration, seeing as fluorescence lifetimes only decreased significantly after 40 minutes, and the average count rate decreased steadily to $<$ 10\% after 60 minutes. This can be explained by a slightly higher protein-to-detergent ratio, which means that more detergent was bound to LHCII trimers. \par

To further elucidate the reduction of LHCII fluorescence during detergent removal, it was necessary to directly quantify changes in the absorption cross-section. According to Eq. \ref{eq:aggregation_condition}, the relative absorption cross-section of LHCII aggregates compared with LHCII trimers can be determined from $\langle N \rangle$. The ensemble-averaged number of LHCII trimers per aggregate during Bio-Bead incubation is given by
\begin{equation}
    M = \frac{\langle N_0 \rangle}{\langle N \rangle} = \frac{C_0}{C},
    \label{eq:aggregate_composition}
\end{equation}

\noindent where $C_0$ is the LHCII concentration before Bio-Beads incubation and $C$ is the concentration after some detergent removal time. Hence, combining Eqs. \ref{eq:aggregation_condition} and \ref{eq:aggregate_composition}, an aggregate of $M$ trimers has an absorption cross-section of $\langle \sigma^{eff} \rangle = M\sigma$. We assumed $\sigma = 1.7 \times 10^{-15}$ cm$^{-2}$ at 633 nm for an individual trimer \cite{Kruger2010}. \par
\begin{figure}[hbt!]
    \centering
    \includegraphics[scale=0.37]{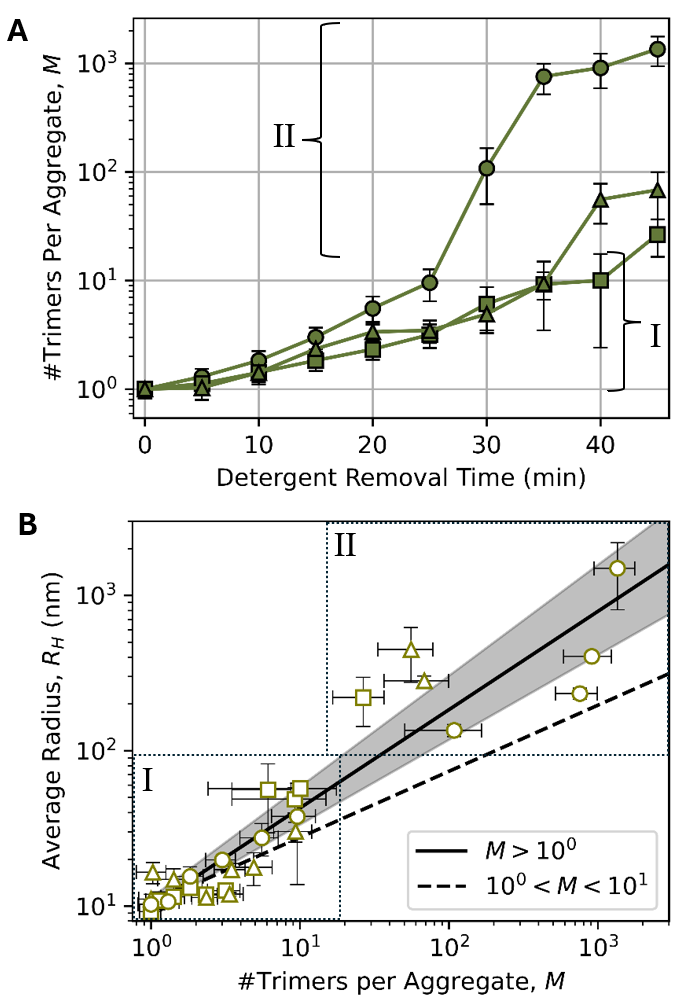}
    \caption{\textbf{A}. Time-dependent increase in the average number of LHCII trimers per aggregate, $M$, calculated through the relative inverse concentration (Eq. \ref{eq:aggregate_composition}), for three repetitions of detergent removal with Bio-Beads. Each shape represents a different experimental run. I and II refer to the slow and fast aggregation phases, respectively. \textbf{B}. Hydrodynamic radius against $M$ for the three repetitions in A (data points) along with a double logarithmic linear regression (solid line) with a corresponding 95\% confidence interval (shaded area). The gradient is 0.63 $\pm$ 0.04, with $r^2$ = 0.88, giving a fractal dimension of $d_f = 1.6 \pm 0.1$. The linear regression for the small aggregates ($0 < M < 10$, dashed line) gives $d_f = 2.3 \pm 0.4$. The dotted outlines indicate the approximate borders between Phases I and II of the aggregation process.}
    \label{fig:R_vs_trimers_WT_LHCII}
\end{figure}

The increase in \textit{M} with detergent removal time is shown in Figure \ref{fig:R_vs_trimers_WT_LHCII}A. A sudden increase in the slope in the region $10 \lesssim M \lesssim 20$ indicates a transition from an initial slow phase (labeled I) to a faster growth phase (labeled II). This biphasic behavior corresponds to the time-dependent changes in the average lifetimes and hydrodynamic radii in Figures \ref{fig:I_L_N_R_WT_LHCII}B and D, which is reminiscent of the two kinetic phases of detergent desorption from membrane proteins reported by Wolfe et al. \cite{Wolfe2018}. When the detergent--protein binding strength is greater than the detergent--detergent binding, then a pre-desolvation phase (I) of detergent proteomicelles below the CMC is apparent, followed by a rapid desolvation phase (II). During phase I, a mixture of LHCII trimers and oligomers resulted in values of $R_H$ between 10 and $\sim$100 nm. During the second phase, most LHCIIs were aggregated and the average aggregate size was markedly greater than in phase I, resulting in $R_H$ values of $\gtrsim$100 nm. \par

The correlation between $R_H$ and $M$ (Fig. \ref{fig:R_vs_trimers_WT_LHCII}B) allows $M$ to be used as an indicator of the average composition of aggregates. Assuming that LHCII aggregation occurred in a fractal fashion, $R_H$ has a power-law relationship with the number of constituents, $M$:
\begin{equation}
    {R_H} =  {R_0}{M}^{1/d_f},
    \label{eq:hydrodynamic_model}
\end{equation}

\noindent where $R_0 =$ 10.2 $\pm$ 0.9 nm is the measured hydrodynamic radius of LHCII trimers and the power-law exponent $\nu = 1/d_f$ is the inverse fractal dimension of the aggregates. The value of $d_f=1.6 \pm 0.1$, obtained from a linear regression of all data points (i.e., $M > 10^0$), corresponds well with the fractal dimension of a diffusion-limited cluster aggregate (DLCA). For small $M$, the data follows a flatter slope, giving $d_f = 2.3 \pm 0.5$ and indicating that either planar aggregates formed or more compact, reaction-limited cluster aggregates (RLCA) occurred \cite{Bunde1992, PhysRevE.103.012138, Jungblut2019}. \par

We note that the measured $R_H$ values can only be directly interpreted as the radius of a sphere with the same translational diffusion time as the LHCII aggregates. The strong deviation between the measured $R_H$ values of large aggregates ($R_H > 100$ nm) and the trend lines in Figure \ref{fig:R_vs_trimers_WT_LHCII}B confirm their significant shape variation, indicating that the exact shape cannot be determined from translational diffusion studies alone. \par

In summary, Figure \ref{fig:R_vs_trimers_WT_LHCII}A demonstrates that partial aggregation started occurring from the onset of Bio-beads incubation and complete aggregation occurred after some time (approximately 30 -- 40 minutes). Figure \ref{fig:R_vs_trimers_WT_LHCII}B demonstrates that aggregation started slowly and was initially characterized by a high fractal dimension, indicating a more ordered orientation-dependent clustering process. Thereafter, after the CMC of the detergent was reached, more rapid diffusion-limited aggregation with a lower fractal dimension occurred. It should be noted that Eq. \ref{eq:hydrodynamic_model} is a phenomenological relation and that the measured hydrodynamic (Stokes) radius is implicitly dependent on the shape of the macromolecules. It is also influenced by the detergent micelles and hydrodynamic shells around LHCII trimers and aggregates. 
%The absolute sizes of proteins can only be obtained with cryo-EM (see \cite{Standfuss2005, Liu2004} for the absolute size of LHCII trimers).
Since the $M$-values obtained were not affected by hydrodynamic properties, it can be regarded as a more convenient measure of the ensemble-averaged sizes of LHCII aggregates. \par 

\section{Modeling}
\label{sec:analytical_modeling}
To interpret the fluorescence kinetics of LHCII aggregation during detergent removal, we employed analytical models of quenching and EEA to describe the observed trends. We found these models surprisingly successful for interpretational purposes. \par

\subsection{Quenching}
\label{subsec:trapping}
The quantum yield of LHCII aggregate fluorescence without any (nonlinear) EEA can be described by
\begin{equation}
    \Phi = \frac{k_r}{k_r + k_{IC} + k_{ISC} + k_Q},
    \label{eq:yield_fluorescence}
\end{equation}

\noindent where $k_r$ is the intrinsic rate of LHCII radiative decay, $k_{IC}$ is the nonradiative internal conversion rate, $k_{ISC}$ is the rate of intersystem crossing of singlet excited Chl states to triplet states in LHCII, and $k_Q$ is the rate of nonradiative energy dissipation due to exciton trapping by quenching centers, assuming a Stern-Volmer process \cite{Ilioaia2008}. The values $k_r+k_{IC}=(5.81$ ns$)^{-1}$ and $k_{ISC}=(8.54$ ns$)^{-1}$ were taken from Ref. \cite{Gruber2015}. Taking $k_r+k_{ISC}=(5.5$ ns$)^{-1}$ from Ref. \cite{Huyer2004}, it can be deduced that $k_{IC} = (9.3$ ns$)^{-1}$ and $k_r= (15.5$ ns$)^{-1}$. From these values, $\Phi\approx0.22$ if no quenching occurs, which corresponds to Ref.~\cite{Palacios2002}. The rate of quenching in an aggregate, $k_Q$, can be assumed to be proportional to the density of quenchers in the aggregate, $\rho_q$, assuming a constant rate of quenching per LHCII trimer, $k_q$, giving $k_Q = k_q \rho_q$. The total linear decay rate of excited states is the reciprocal of the average fluorescence lifetime, $k_f =1/\tau^{avg} = k_r + k_{IC} + k_{ISC} + k_q \rho_q$. Excitation quenching ($Q$) becomes more probable as a result of increases in $\rho_q$, which causes observable decreases in the absolute fluorescence yield measurable with TCSPC. This was done by resolving the 200 -- 600 ps fluorescence decay component that increased in amplitude as $M$ or $R_H$ increased (see Figs. \ref{fig:I_L_N_R_WT_LHCII}E, F, G and \ref{fig:ps_L_STA_WT_LHCII}). Based on Eq. \ref{eq:yield_fluorescence}, the relative fluorescence yield subject only to quenching (i.e., excluding EEA) is
\begin{equation}
    \frac{\Phi}{\Phi_0} = \frac{\tau^{avg}}{\tau^{avg}_0} = \frac{1}{1 + \tau^{avg}_0 k_Q},
    \label{eq:Q}
\end{equation} 
\noindent where $\Phi_0$ is the intrinsic fluorescence yield of LHCII and $\tau^{avg}_0 = (k_r + k_{IC} + k_{ISC})^{-1} = 3.6$ ns is the fluorescence lifetime of LHCII trimers without quenching. The amplitude-averaged fluorescence lifetimes can be used to calculate the singlet excited-state quenching constants $K_D = \tau^{avg}_0 k_Q$. We note that Eq. \ref{eq:Q} can be rewritten as the Stern-Volmer relation for dynamic quenching: $K_D = \tau^{avg}_0/\tau^{avg} - 1$.

The fraction of singlet excitons that are quenched in an LHCII aggregate can be interpreted as the average fraction of trimers in an aggregate that contains quenching traps, which is
\begin{equation}
    Q = \frac{K_D}{1+K_D} = 1-\frac{\tau^{avg}}{\tau^{avg}_0}.
    \label{eq:quenched_fraction}
\end{equation}
\par

\subsection{SSA}
\label{subsec:SSA}
In the case where nonlinear or bimolecular excitonic processes (i.e., EEA) in LHCII aggregates are significant, the effective fluorescence yield, $\Phi^{eff}$, depends on SSA and STA in addition to quenching. This can be approximated by identifying the process(es) that contributed most to the decrease in fluorescence count rates during aggregation. \par

The instantaneous probability that an excitation is annihilated by another excitation is determined by the number of singlets and triplets present in an LHCII trimer or aggregate at an instant in time. This depends on the average intensity of an excitation pulse. The Pauli master equation approach can be used to obtain approximations for $\Phi^{eff}$ based on a three-level model \cite{Barzda2001, Gruber2015}, known as the statistical approach. However, for steady-state fluorescence yields, a time-independent approach is appropriate. In this case, when the measured relative fluorescence yields at increasing excitation rates correspond to the shape of an approximated analytical formula, the type of annihilation involved can be identified \cite{Paillotin1979, Paillotin1983}. This is the approach used in this section. \par
The average number of excitations generated per pulse, $n_0$, can be calculated from the average pulse intensities, $J_e = \frac{1}{T}\int^T_0 J(t) dt \approx I_e \tau_{\Delta}$, where $\tau_{\Delta}$ is the time between consecutive pulses (i.e., 50 ns in this study). This allows for approximating the average number of excitations as $n_0 = \sigma J_e$. Based on this, the Poisson-distributed probability that two or more excitations were generated in isolated LHCII trimers at $<$ 20 $\mu$W laser power was calculated to be $<0.1\%$, rendering SSA negligible prior to significant aggregation. \par
When the diffusion-limited rate constants of SSA and STA, $\gamma_{SS}$ and $\gamma_{ST}$, are much higher than $k_f =1/\tau^{avg}$, $\gamma \tau^{avg}>>1$ and the ``fast annihilation" regime applies. This warrants a statistical approach to calculate pulse-integrated fluorescence yields. In contrast to this, the ``extended aggregates" regime is applicable when $\gamma \tau^{avg}\approx 1$, allowing a time-independent approach to calculate instantaneous fluorescence yields \cite{Barzda2001}. When only steady-state fluorescence is considered, however, instantaneous fluorescence yields are sufficient for both STA regimes in LHCII aggregates \cite{Paillotin1983}. \par

The SSA rate in LHCII aggregates normalized to trimers has been determined to range between (16 ps)$^{-1}$ and (24 ps$)^{-1}$ \cite{Barzda2001} and the STA rate in single LHCII trimers has been calculated as $\gamma_{ST} = (35 $ ps$)^{-1}$ \cite{Gruber2015}. These rate constants are two orders of magnitude higher than the intrinsic fluorescence rate of LHCII ($k=1/\tau_0 =$ ($3.6$ ns)$^{-1}$). Assuming that the annihilation rates are linearly scaled down by the number of sites per aggregate to which excitons can randomly diffuse, the ``fast annihilation" regime is expected to be valid for LHCII trimers and small aggregates (i.e., $1 \le M \lesssim 100$).
The Poissonian time-integrated fluorescence yield can then be used to determine the effective yield due to the influence of SSA \cite{Barzda2001, Paillotin1979, vanGrondelle1985, Bernhardt1999}:
\begin{equation}
    \mathrm{SSA} : \Phi^{eff} = \Phi \frac{1 - e^{-n(M)}}{n(M)},
    \label{eq:SSA_FL_yield}
\end{equation}

\noindent where $n$ is the average number of Chl excitations per LHCII aggregate of $M$ trimers generated by a laser pulse, given by
\begin{equation}
    n(M) = \sigma^{eff}J_e = \sigma M J_e,
    \label{eq:singlet_states_per_pulse}
\end{equation}

\noindent assuming that aggregation increases the absorption cross-section linearly and the average pulse intensity is constant. For our FCS measurements on trimers ($M = 1$), $n_0 = 0.05$, based on the average photon density per pulse of $J_e = 4.3 \times 10^{13}$ photons$\cdot$cm$^{-2}$. Thus, having $n_0 << 1$, SSA in LHCII trimers was very improbable during the FCS measurements. \par

\subsection{STA}
\label{subsec:STA}
The effect of STA on the fluorescence yield of LHCII depends on the number of Car triplet states with which a singlet exciton can annihilate. Given that $k_{ISC}$ for Chl in LHCII remains constant during aggregation \cite{Barzda1998} and that the triplet--triplet transfer from Chls to Cars can be assumed to occur more than 90\% of the time \cite{Peterman1995, Barzda1998}, a ``trap-limited" model can be used for STA. In such a model, all Chl triplet states are generated at the same rate and assumed to be transferred to a nearby Car, rendering these resultant triplet excitons stationary, and thus the rate of STA is only singlet diffusion-limited \cite{vanGrondelle1985, Gruber2015}. Since these Car triplets are long-lived compared to the time between excitation pulses, they quickly accumulate, leading to large steady-state populations and increased probabilities of STA. Thus, STA is expected to be a significant cause of reduced fluorescence yields, even at moderate excitation pulse intensities \cite{Valkunas1991}. A Stern--Volmer-type relation has been shown to be a sufficient steady-state approximation for large photosynthetic units and single LHCII trimers, meaning that the fast annihilation regime applies \cite{vanGrondelle1985, Breton1979}. Incorporating the saturation effect of the triplet states due to STA can thus be done with a Stern-Volmer interpretation, constrained by the number of Cars per Chl in LHCII. Building on Refs. \cite{Paillotin1983, Valkunas1991}, the instantaneous effective fluorescence yield subject to STA after reaching the steady state can be approximated by:
\begin{equation}
     \mathrm{STA}: \Phi^{eff} = \dfrac{\Phi}{1 + \alpha\gamma_{ST}\tau^{avg}(M) N_T(M)},
    \label{eq:STA_FL_yield}
\end{equation}

\noindent where $N_T$ is the steady-state Car triplet population per domain in an aggregate of $M$ trimers (assuming the aggregate is not significantly larger than the domain size), $\alpha$ is the density of Cars in LHCII that contribute to the triplet population, and $\gamma_{ST}\tau^{avg}=k_{ST}$ is the Stern--Volmer constant for STA in an LHCII trimer. \par
The applicability of both Eqs. \ref{eq:SSA_FL_yield} and \ref{eq:STA_FL_yield} was especially aimed at explaining the observed decrease in fluorescence yield before $\tau^{avg}$ decreased significantly during detergent removal, which can be assumed to involve the formation of small aggregates (see Section \ref{sec:results}, in particular with reference to Figs.  \ref{fig:I_L_N_R_WT_LHCII} -- \ref{fig:R_vs_trimers_WT_LHCII}). 

To estimate the values of $N_T$ for LHCII trimers and aggregates in the absence of SSA, which is accurate when SSA is much less probable than STA \cite{Zaushitsyn2007}, the following expression can be used to describe the accumulation of triplets to a steady-state population under pulsed excitation:
\begin{equation}
    N_T(M) = \Phi_T \frac{\sigma I_e M}{K_t \tau_{\Delta}},
    \label{eq:N_M}
\end{equation}
\noindent where $\Phi_t$ is the triplet yield in LHCII, $M$ is again the number of LHCII trimers per aggregate, and $K_t={\tau_{car}}^{-1}=2.5\times10^5$ s$^{-1}$ is the decay rate of Car triplet states in LHCII, assuming $\tau_{car}=4\ \mu $s. Eq. \ref{eq:N_M} is based on the assumption that the domain size of a diffusing singlet excitation spans the entire aggregate. The triplet yield is rate-dependently decreased by STA in a larger aggregate (due to the quenching of a singlet excitation that could otherwise undergo intersystem crossing) and can be written as
\begin{equation}
    \Phi_T = \frac{k_{ISC}}{k_f+\gamma_{ST}N_T/M},
    \label{eq:triplet_yield}
\end{equation}

\noindent assuming that $\gamma_{ST}$ is scaled down by $M$. Rearranging Eq. \ref{eq:N_M} and substituting Eq. \ref{eq:triplet_yield} then gives %\cite{Gruber2015, Zaushitsyn2007}
\begin{equation}
    N_T(M) = \frac{-k_f + \sqrt{k_f^2 + 4\gamma_{ST}k_{ISC}\tau_{car}\sigma J_e/\tau_{\Delta}}}{2 \gamma_{ST}}M = N_{T_0} M,
    \label{eq:analytical_triplets}
\end{equation}

\noindent where $N_{T_0}$ is the predicted triplet population in an LHCII trimer independent of aggregate size, which was calculated to be constant at 0.13 for $n_0=\sigma J_e = 0.05$. This approximates the density of triplets in an LHCII aggregate in terms of the triplet population divided by the number of trimers per aggregate, $M$, assuming that the domain size of exciton diffusion in the aggregate spans the whole aggregate. \par

\subsection{Kinetic Model}
\label{subsec:kinetic_modeling}
We next considered a non-stochastic kinetic model to combine STA, SSA, and Q in LHCII aggregates, scaled by $M$, to calculate the steady-state triplet populations. A deterministic framework was chosen because aggregate sizes can be readily incorporated via the $M$-values. Furthermore, explicit calculation of fluorescence lifetimes was unnecessary, as normalized fluorescence yields calculated using a steady-state approximation (Eq. \ref{eq:STA_FL_yield}) were found to agree well with the measured relative fluorescence intensities. The effect of stimulated emission was assumed to be negligible. Following Barzda et al. \cite{Barzda2001} and Gruber et al. \cite{Gruber2015}, the singlet ($n$) and triplet ($N$) populations evolve according to the following kinetic models:
\begin{equation}
    \frac{dn}{dt} = \sigma J_e M - \frac{n}{\tau^{avg}} - \frac{\gamma_{ST}}{M}Nn - \frac{1}{2}\frac{\gamma_{SS}}{M}n^2 ,
    \label{eq:singlet_equation}
\end{equation}

\begin{equation}
    \frac{dN}{dt} = k_{ISC}n - K_tN.
    \label{eq:triplet_equation}
\end{equation}

Because $n$ and $N$ represent the total singlet and triplet populations within an aggregate, the annihilation rate constants ($\gamma_{SS}$ and $\gamma_{ST}$) were scaled down by $M$ to account for the effect of limited exciton diffusion domains in larger aggregates, as originally proposed by Paillotin et al. \cite{Paillotin1979}. Steady-state triplet populations for small aggregates were then used to assess the accuracy of the kinetic model, replacing Eq. \ref{eq:analytical_triplets}, by using Eq. \ref{eq:STA_FL_yield} to calculate the resultant effective fluorescence yields. Simulations under pulsed excitation were performed by solving the kinetic model separately for each pulse over 400 pulses, whereas simulations under continuous excitation were performed by solving the model directly over a 20-$\mu$s interval. Even for pulsed excitation, the logistic form of Eq. \ref{eq:triplet_equation} resulted in steady-state triplet populations being reached after around 10 $\mu$s, corresponding to approximately 200 pulses. To compare the simulation with fluorescence data, a range of small aggregates was considered ($1 \le M < 10$) and the steady-state triplet populations were assumed to be $N_T \approx N(t = 20 \ \mu\text{s})$. \par

\section{Modeling Results}
\label{sec:modeling_results}

\begin{figure*}[hbt!]
    \centering
    \includegraphics[scale=0.30]{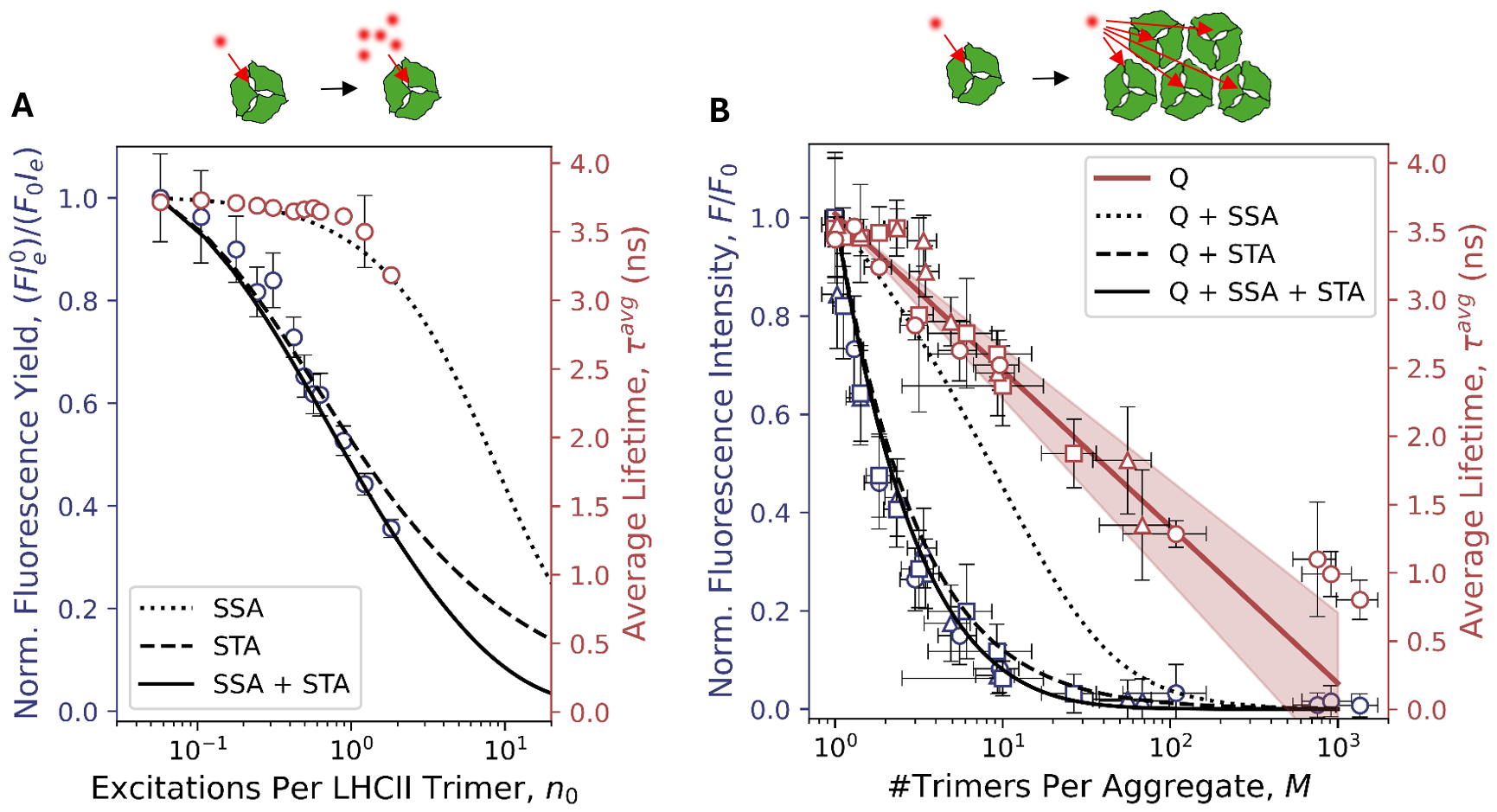}%width=\linewidth
    \caption{ \textbf{A}. Relative fluorescence yield of LHCII trimers (blue circles) at increasing excitation intensities on a logarithmic scale, effectively increasing $n_0$ from 0.05 to 5 for $M=1$, calculated theoretical fluorescence yield taking into account only SSA (dotted line, Eq. \ref{eq:SSA_FL_yield}), only STA (dashed line, Eq. \ref{eq:STA_FL_yield}), and both SSA and STA (solid line), and amplitude-averaged fluorescence lifetimes (red circles).
    \textbf{B}. Relative fluorescence intensity ($F/F_0$, blue symbols), equivalent to the relative effective yield of fluorescence after incubation with Bio-Beads started ($\Phi^{eff}/\Phi^{eff}_0$), and average two-component fluorescence lifetime ($\tau^{avg}$, red symbols), displayed against $M$. A linear regression fit of $\tau^{avg}$ (red line) approximates $\Phi/\Phi_0$ based on quenching, Q, giving a gradient of $-0.49 \pm$ 0.03 ns/ln(M) and an $r^2$-coefficient of 0.94. Calculated $\Phi^{eff}/\Phi^{eff}_0$ based on SSA (Eq. \ref{eq:SSA_FL_yield}) and Q (Eq. \ref{eq:Q} via the $\Phi/\Phi_0$ regression fit) (dotted line), based on STA (Eq. \ref{eq:STA_FL_yield}) and Q (dashed line), and Q, SSA, and STA combined (solid line).}
    \label{fig:norm_yields__lifetimes__vs_J_M}
\end{figure*}

Eqs. \ref{eq:SSA_FL_yield} and \ref{eq:STA_FL_yield} show how the effective fluorescence yield of LHCII during aggregation is expected to depend on SSA and STA. However, it was not clear how the measured fluorescence lifetimes depend on fast annihilation processes that occur faster than the excitation pulse duration and IRF. To investigate the effect of annihilation on fluorescence lifetimes for the measurement setup, LHCII trimers were excited at increasing intensities starting at the intensity used for FCS, $I^0_e = 270$ W$\cdot$cm$^2$. Figure \ref{fig:norm_yields__lifetimes__vs_J_M}A shows measured relative effective fluorescence yields based on average intensities ($(FI^0_e)/(F_0I_e) = \Phi^{eff}/\Phi^{eff}_0$, following Eqs. \ref{eq:countrate} and \ref{eq:aggregation_condition}) of LHCII trimers with increasing excitation intensity, normalized to the fluorescence count rate for 0.05 excitations per trimer at $I_e^0$. \par
The lack of correspondence between the measured relative intensities (blue circles) and Eq. \ref{eq:SSA_FL_yield} (dotted curve) shows that SSA was not the main cause of the reduction in steady-state fluorescence yield with increasing excitation. Notably, the average fluorescence lifetime (red circles) started to be affected by annihilation when $n_0 \approx 1$ and its dependence on $n_0$ corresponds well with the trend predicted by Eq. \ref{eq:SSA_FL_yield}, suggesting that SSA is the dominant cause of the reduction in lifetime at high excitation densities. However, it is clear that $\tau^{avg}$ was not affected by SSA or STA as significantly as the average fluorescence intensities. \par

While the effect of SSA alone strongly overestimates the relative fluorescence yields of LHCII trimers (Fig. \ref{fig:norm_yields__lifetimes__vs_J_M}A, dotted line), the effect of STA, as determined through Eqs. \ref{eq:STA_FL_yield} and \ref{eq:analytical_triplets}, predicts fluorescence yields well for small to moderate excitation intensities (Fig. \ref{fig:norm_yields__lifetimes__vs_J_M}A, dashed line), while the combined effect of SSA and STA (Fig. \ref{fig:norm_yields__lifetimes__vs_J_M}A, solid line) reproduces the experimental values very well. It should be noted that Figure \ref{fig:norm_yields__lifetimes__vs_J_M} does not show fits but a direct comparison between the predicted and measured behavior. In our calculations, we considered 12 Cars per LHCII trimer with 42 Chls, yielding a quencher density of $\alpha = 3.5$, which was assumed to be the same across all measurements. 
\par

The fluorescence yields of LHCII during aggregation at a constant excitation intensity ($I_e^0$) were calculated using the same expressions for SSA and STA, but now with $n_0$ given by Eq. \Ref{eq:singlet_states_per_pulse} depending explicitly on $M$. To account for quenching in the calculations, the effect of aggregation-related quenching (Q) was incorporated into the linear decay rate (see Ref. \cite{Rutkauskas2012}) by the fit of $\tau^{avg}$ vs. $\ln(M)$ (Fig. \ref{fig:norm_yields__lifetimes__vs_J_M}B, red line). This approximates the effect of quenching on the fluorescence yield, independent of annihilation, provided that SSA and STA had minimal effect on $\tau^{avg}$ for LHCII trimers at increasing excitation rates, and that this remains true when the absorption cross-section increases during aggregation. \par
The effect of Q alone (red line) significantly underestimates the strong reduction in the experimental fluorescence yields (blue data) with increasing aggregate size, because of the large deviation between the decreases in $F/F_0$ and $\tau^{avg}$. Adding SSA (dotted line), the resultant calculated fluorescence yield curve still deviated substantially from the experimental data, while STA, in addition to Q (dashed line), results in a curve that is very close to the experimental values. The combination of all three size-dependent quenching processes (solid line) reproduces the experimental fluorescence yield remarkably well. Again, except for Q, the experimental data were not fitted but overlaid with the predicted trends based one simple, time-independent theoretical approximations.

The comparison of relative fluorescence intensities with the calculated relative fluorescence yields shows that increasing the absorption cross-section due to LHCII aggregation leads to increased interaction between triplet excitons and diffusing singlet excitons, resulting in increased STA, similar to the effect of higher excitation rates in LHCII trimers, without a measurable decrease in fluorescence lifetimes. 
Notably, average fluorescence lifetimes below 1 ns corresponded to very large aggregates ($M>100$), even with both quenching and STA involved, but relative fluorescence intensities below 0.1 were apparent already for much smaller aggregates ($M \approx 10$). \par

Assuming that the fast annihilation condition was met for small aggregates, excitation annihilation occurred within the duration of the excitation pulse and was much faster than the average experimental fluorescence decay times.
%(see Fig. \ref{fig:LHCII_decay_and_autocorrelation_dynamics}B). 
Based on Figure \ref{fig:norm_yields__lifetimes__vs_J_M}B, STA was the main contributor to the decreases in fluorescence yield in LHCII aggregates, while the measured fluorescence lifetimes decreased from 3.6 to 2.5 ns for $M\le10$ (i.e., for small aggregates). The measured relative effective fluorescence count rates and lifetimes, $F/F_0$ and $\tau^{avg}/\tau^{avg}_0$, were used to empirically estimate the average triplet population per aggregate, $N_T$, if the STA domain stretches over a whole aggregate. By assuming a Stern--Volmer interpretation for STA, rearranging Eq. \ref{eq:STA_FL_yield} gives the average steady-state triplet population in an aggregate as
\begin{equation}
    N_T \approx \frac{F_0}{F} \left[\frac{1}{\tau^{avg}_0\gamma_{ST}\alpha}+N_{T_0}\right] - \frac{1}{\tau^{avg}\gamma_{ST}\alpha},
    \label{eq:experimental_steady_triplets}
\end{equation}

\noindent where $\tau^{avg}_0$ is the maximum average measured fluorescence lifetime of trimers ($M=1$). This approximation assumes that the aggregates are smaller than the exciton-diffusion domain sizes. The corresponding triplet fraction, defined as the fraction of LHCII trimers in an aggregate of $M$ trimers that contains a Car triplet after reaching steady state, is 
\begin{equation}
    T = \frac{N_T}{M},
    \label{eq:triplet_fraction}
\end{equation}
\par

\noindent which is an experimentally acquired quantity assumed to depend on the aggregate size, while $N_{T_0}$ is the triplet population in a trimer that was theoretically calculated for $M=1$ with Eq. \ref{eq:analytical_triplets}. \par

\begin{figure}[hbt!]
    \centering
    \includegraphics[scale=0.65]{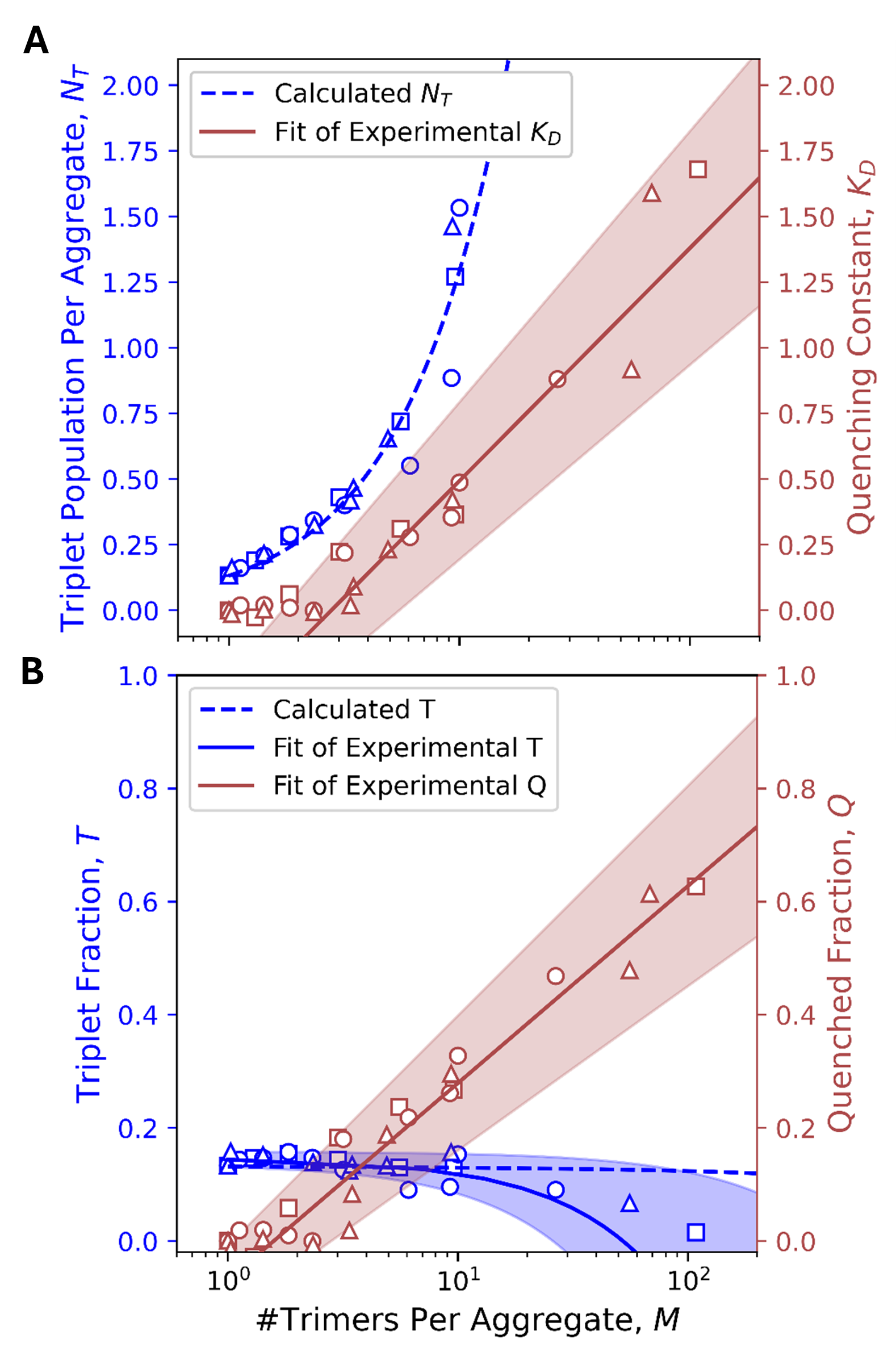}
    \caption{A. Number of Car triplet states per LHCII aggregate as a function of the aggregate size $M$, determined from Eq. \ref{eq:experimental_steady_triplets} (blue data) using the experimental $F/F_0$ and $\tau^{avg}$ values measured with SPAD 1 at 270 $\pm$ 30 W$\cdot$cm$^{-2}$ excitation intensity, and calculated based on the kinetic model through Eqs. \ref{eq:singlet_equation} and \ref{eq:triplet_equation} (blue dashed line, Calculated $N_T$). Quenching constant $K_D$ as a function of the aggregate size $M$ (red data), obtained through the experimental $\tau_{avg}$ values (Eq. \ref{eq:quenched_fraction}), and a linear-log fit of the form $K_D=m_K \ln{(M)}+b_K$ (Fit of Experimental $K_D$, red line) with the 95\% confidence region shaded in red. See Table \ref{tab:fitting_results} for the linear regression results. B. Triplet fractions, $T$, corresponding to the $N_T$ values in A., determined from Eq. \ref{eq:triplet_fraction} and shown as a function of the aggregate size $M$ (blue data). A linear fit ($T=m_T M + b_T$) on a linear scale for datapoints with small errors, i.e. for $1\le M < 20$, is shown as the blue solid line (Fit of Experimental T), with the 95\% confidence region as a blue shaded area. Triplet fractions corresponding to the populations in A., calculated from Eq. \ref{eq:triplet_fraction}, are shown as a blue dashed line (Calculated T). Fraction of LHCII trimers per aggregate that are quenched, determined from the same $\tau^{avg}$ data using Eq. \ref{eq:quenched_fraction} (red data) and linear-log fit of the form $Q=m_Q \ln{(M)}+b_Q$ (Fit of Experimental Q, red line) with the 95\% confidence region shaded in red. See Table \ref{tab:fitting_results} for the fitting results.} 
    \label{fig:comparisons_of_populations}
\end{figure}

The validity of Eq. \ref{eq:experimental_steady_triplets} depends on the assumption that $\tau^{avg}$ was minimally affected by STA and SSA during the initial stage of aggregation (phase I, $1\le M < 10$---see Fig. \Ref{fig:R_vs_trimers_WT_LHCII}), as is the case for LHCII trimers when $0.05 \le n_0 < 0.5$ (see Fig. \ref{fig:norm_yields__lifetimes__vs_J_M}A). This suggests that $\Phi / {\Phi}_0 \approx \tau^{avg} / \tau^{avg}_0$ for small aggregates, equivalent to the case for trimers (see Eq. \ref{eq:Q}). The Chl-to-Car ratio, $\alpha$, was kept constant at $3.5$. \par

Figure \ref{fig:comparisons_of_populations}A shows how the experimentally estimated triplet populations  of the LHCII aggregates, according to Eq. \ref{eq:experimental_steady_triplets}, increases with increasing aggregate size, $M$, considering a constant excitation intensity and $M < 200$. A deviation of the triplet populations from quenching constants calculated from the average lifetimes is clear, suggesting that triplets and quenching traps are distinct. The values calculated using the phenomenological kinetic model (Eqs. \ref{eq:singlet_equation} and \ref{eq:triplet_equation}), shown as the blue dashed line, closely follow the data points for $1\le M \le 10$, suggesting that the kinetic model is also accurate for small aggregates. \par
The NPQ quenching constant, $K_D$, calculated from the average lifetimes in Eq. \ref{eq:quenched_fraction}, shows a semi-logarithmic dependence on $M$, indicating that the number of energy traps that cause quenching increases by large increases in aggregate size. The linear fit indicates that $K_D\sim1$ at $M\sim35$, i.e., a 50\% lifetime reduction requires at least 35 interconnected trimers. This aggregate size can be considered the functional domain size of LHCII aggregates that achieve a quenching level equivalent to a minimum NPQ state.

\begin{table}[hbt!]
    \centering
       \caption{Linear regression results of the fits in Fig. \ref{fig:comparisons_of_populations} for the different measures of energy traps and triplets vs. $M$.}
    \begin{tabular}{c|c|c|c}
          & Gradient & Intercept & r$^2$ \\
         \hline
        $K_D$ & 0.39 $\pm 0.03$ & $-0.39 \pm$ 0.08 & 0.93 \\
        $T$ & $-0.003 \pm$ 0.001 & 0.15 $\pm$ 0.01 & 0.92 \\
        $Q$ & 0.15 $\pm$ 0.02 & $-0.07 \pm$ 0.03 & 0.94 
    \end{tabular}
     \label{tab:fitting_results}
\end{table}

Figure \ref{fig:comparisons_of_populations}B indicates that experimental triplet fractions calculated from the $N_T$ values using Eq. \ref{eq:triplet_fraction} also show a distinctly different dependence on $M$ than the quenched fraction (Eq. \ref{eq:quenched_fraction}). The correspondence of the theoretically calculated $T$ values (dashed blue line) exemplifies that the simple kinetic model is sufficiently accurate for steady-state calculations. The linear fit of $T$ vs. $M$ (solid blue line) gives a near-constant triplet fraction for $1\le M \le 10$, with $T=0.15$ at $M=1$, very close to $T_0=0.13$ predicted by Eq. \ref{eq:analytical_triplets}. Experimental triplet populations in aggregates comprising 20 or more trimers ($M \ge 20$) could not accurately be determined from the $F/F_0$ measurements, owing to the low concentrations of the highly aggregated samples. However, a deviation from the theoretical prediction is apparent in Figure \ref{fig:comparisons_of_populations}B, which suggests an extended aggregate regime wherein STA is saturated by domain-limited exciton diffusion, i.e., not all triplet excitons can be interacted with by diffusing singlet excitons in large aggregates. \par

\section{Discussion}
\label{sec:discussion}

In this work, small LHCII ensembles during aggregation were investigated to link particle size to changes in fluorescence kinetics. By combining FCS with TCSPC on isolated LHCII in aqueous solution while incrementally lowering the detergent concentration, the gradual formation of aggregates could be monitored through diffusion-based hydrodynamic radii, while fluorescence intensities and lifetimes were recorded simultaneously. In this way, increasing aggregate radii and compositions could be directly correlated with decreasing fluorescence yield and shortening of the amplitude-averaged lifetime, embedding the observed fluorescence kinetics within a quantitatively defined structural context.

The methodology revealed a consistent relationship between hydrodynamic radii inferred from diffusion times and the number of LHCII trimers per aggregate, estimated from the autocorrelation amplitude. Thus, the number of particles detected in the confocal volume could be interpreted as a scaling factor for the effective absorption cross-section as aggregation proceeded. Within this structural framework, a persistent nonlinear relationship between fluorescence count rates and amplitude-averaged fluorescence lifetimes emerged as detergent was removed and aggregate size consequently increased, indicating that distinct photophysical processes modulate these quantities during aggregation. This trend was reproduced for LHCII trimers across different excitation intensities.

Although our results give clear evidence that this observation can be well-explained by enhanced EEA arising from increased pigment interconnectivity within aggregates, it is worth investigating two alternative possible mechanisms: (1) pigment loss from LHCII due to detergent adsorption to Bio-Beads and (2) adsorption of LHCII onto Bio-Beads. The first explanation is unlikely in light of circular dichroism measurements of LHCII aggregates prepared with Bio-Beads, which showed no significant alteration of pigment–protein interactions \cite{Akhtar2015}. The second is also improbable, since detergent removal with Bio-Beads typically yields near-complete protein recovery \cite{Horigome1983, Varhac2009}.

To minimize annihilation effects, the TCSPC-FCS measurements were performed at excitation intensities that are particularly low for FCS. Nevertheless, excitation annihilation, in particular STA, contributed significantly to the decrease in fluorescence yield. This was evident from the agreement between the measured relative fluorescence yields and a time-independent STA-based model of the effective fluorescence yield, whereas an SSA-based model was insufficient. However, the corresponding $\tau^{avg}$ values did not follow the same trend. The apparent discrepancy reflects the different ways in which annihilation processes influence steady-state intensities and pulse-integrated lifetimes. Specifically, for LHCII aggregates excited with laser pulses of tens of ps duration at MHz repetition rates, most SSA and STA events occur within the excitation pulse duration. Consequently, their effects are partially masked in TCSPC decays by pulse integration and convolution with the IRF. Steady-state fluorescence intensities, by contrast, are directly proportional to the fraction of excitons that ultimately decay radiatively once the steady state is established. STA therefore produces an immediate linear decrease in fluorescence count rates, whereas its influence on $\tau^{avg}$ is nonlinear because TCSPC integrates over excitation pulses whose full width at half maximum exceeds the fastest STA-related decay components. In this sense, the divergence between $F/F_0$ and $\tau^{avg}/\tau^{avg}_0$ during aggregation arises naturally from the different observables, viz., instantaneous yield vs. pulse-integrated decay (see Refs. \cite{Paillotin1979, Paillotin1983, Barzda2001, Gray2024}).

Our analytical approximations of the steady-state triplet population also explained the varying dependence of the fluorescence yield and lifetime on the aggregate size, showing that as the aggregate size increases, the triplet population, and therefore also STA, increases, which lowers the effective fluorescence yield without necessarily producing a corresponding reduction in $\tau^{avg}$. A simple kinetic model predicted that carotenoid triplet populations reach a steady state after a few hundred pulses. The increases in experimentally estimated steady-state triplet populations were well reproduced by the kinetic model. Triplet fractions were largely independent of aggregate size over the range of 1 –- 20 trimers per aggregate, consistent with a scenario in which the triplet population per some fixed number of LHCII trimers within an aggregate remains approximately constant at constant excitation intensity while network connectivity increases. Under these conditions, STA becomes the dominant mechanism reducing fluorescence yields during early aggregation. For aggregates larger than $\sim$10 trimers, the decrease in the triplet fraction becomes significant, possibly reflecting increased quenching rates that compete with intersystem crossing, consistent with a Stern–Volmer interpretation.

The dependence of $\tau^{avg}$ on aggregate size further supports the emergence of additional quenching pathways beyond annihilation. In our combined TCSPC-FCS measurements with a detector exhibiting an IRF of a few hundred ps, the fluorescence decays were dominated by a 200 –- 600 ps component that became more prominent as aggregates grew. This component is reminiscent of singlet excitons migrating to a broad distribution of relatively slow quenching traps within aggregates \cite{Gray2024}. The resulting decays exhibited a biexponential character, and $\tau^{avg}$ decreased approximately logarithmically with increasing $M$ or $R_H$. This trend can be interpreted as reflecting an increasing probability that excitons encounter quenching sites as the pigment network expands. A quantitative interpretation, however, requires simulations that explicitly incorporate pulse duration, repetition rate, and IRF convolution in order to disentangle intrinsic quenching from measurement artifacts. In addition, a stochastic protocol is required to allow each diffusing singlet exciton to interact dynamically with either a quencher or a carotenoid triplet during inter-trimer hopping. Inclusion of effects analogous to entropic repulsion—--which reduce the probability of hopping to a trimer already containing a Car triplet—--may further refine such models and provide a more unified description of annihilation and quenching in aggregated LHCII.

Bulk TCSPC measurements with a faster detector (IRF $\sim$40 ps) resolved additional complexity. Three lifetime components were observed during aggregation, including a 30 –- 90 ps component whose amplitude increased, and a 300 –- 400 ps component associated with quenching. Simulations of multiexponential decays of LHCII aggregates suggest that the reduction in $\tau^{avg}$ reflects at least two distinct quenching mechanisms, giving rise to the appearance of a $<$1 ns component and a reduction of the intrinsic Chl \textit{a} lifetime to approximately 2.5 ns, consistent with previous experimental observations \cite{Johnson2009alteration, Natali2016, Adams2018}. The shortening of the longer-lived component was apparent in the bulk measurements but not in the TCSPC-FCS data with the slower detector, underscoring the importance of temporal resolution. Although the influence of the excitation pulse width on the fastest STA component is dominant, annihilation cannot be excluded as a contributing factor to the reduction of the 3.6 ns component under ps pulsed excitation.

Taken together, these results indicate that LHCII aggregation induces a progressive restructuring of the excitonic landscape, in which increased connectivity enhances both SSA and STA and access to distributed quenching traps. In most transient absorption spectroscopy studies, STA is negligible due to much lower pulse repetition rates, making SSA the main consideration \cite{Barzda2001, Ruban2007, Muller2010}.
However, given the high pulse repetition rates used in time-resolved fluorescence methods such as TCSPC and Streak camera-based spectroscopy, STA becomes the dominant process even at moderate excitation intensities. In addition, the nonlinear divergence between steady-state intensities and time-resolved lifetimes arises naturally in typical TCSPC measurements that use pulse durations of tens to hundreds of ps. It is also the case for continuous illumination when the Poissonian probability that more than one photon is absorbed into an aggregate during the Car triplet lifetime is as high as pulsed excitation with a $>$ 100 kHz repetition rate.

\section{Conclusion}
\label{sec:conclusion}
Performing TCSPC-FCS on LHCII aggregates using MHz-pulsed excitation leads to an accumulation of carotenoid triplet states at intensities often considered to effectively be annihilation-free conditions ($<300$ W$\cdot$cm$^{-2}$, $<10^{21}$ photons$\cdot$cm$^{-2}$, $\le 0.1$ pJ per pulse, $\le 0.05$ excitations per pulse). These conditions cause STA to increase significantly during aggregation, exhibiting much greater sensitivity to the aggregate size than NPQ-related processes at moderate excitation intensities and high pulse rates. This indicates that the rate and extent of NPQ-related quenching in aggregates can easily be overestimated when considering only steady-state fluorescence kinetics, while the role of STA can similarly be underestimated in time-resolved fluorescence measurements. Achieving annihilation-free conditions for time-resolved fluorescence studies, therefore, depends more on STA than SSA. This has often been overlooked in such studies, necessitating a re-evaluation of interpretations of fluorescence kinetics and quenching degrees.

STA causes significant discrepancies between steady-state and time-resolved fluorescence during aggregation at moderate excitation intensities due to most annihilation events occurring during ps excitation pulses and triplet exciton accumulation. Using excitation pulses of a few tens of ps causes STA to have little effect on the measured fluorescence lifetimes, but it is the main cause of the decrease in fluorescence intensities during aggregation.
    
By simultaneously measuring hydrodynamic radii and relative absorption cross-sections with fluorescence kinetics measurements with ps-pulsed excitation, it can be ascertained that excitation quenching due to aggregation is the main cause of observed fluorescence lifetime decreases, even when significant STA occurs, but large aggregates in solution are required for $<$1 ns average lifetimes.

Low-intensity FCS showed a biphasic increase in the hydrodynamic radii of LHCII during detergent removal, implying that the samples contained a distribution of LHCII trimers and aggregates that changed continuously during detergent removal. Relative particle concentrations were found to be sensitive measures of the average aggregate composition of an ensemble of LHCII trimers and aggregates in solution, and this corresponds directly to reductions in fluorescence yield. 

It can be concluded that for fluorescence-based measurements performed with moderate excitation intensities, STA is a major cause of reductions in LHCII fluorescence yield that must be distinguished sufficiently from NPQ-related quenching. Previous TCSPC studies that considered only SSA and quenching may not have correctly interpreted decreases in fluorescence intensity or lifetime because STA may have been underestimated. Due to the sensitivity of the steady-state fluorescence intensity on STA, it would be more reliable to base NPQ-related parameters, such as the quenching constant $K_D$, on the fluorescence lifetime, even when STA is not resolved. However, this requires the use of sub-ps pulses to limit the obscuring effects of annihilation during an excitation pulse.
Further investigation into the interplay between annihilation by Car triplets and quenching by slow energy traps can provide insights into the site domains and densities of energy traps in LHCII aggregates. 

Using femtosecond-pulsed excitation for TCSPC is recommended, as it enables comparison of fluorescence kinetics with stochastic models and time-resolved absorption studies, although this is often not implemented in fluorescence imaging or spectroscopy systems. However, this study demonstrates that, by incrementally aggregating LHCII, the type of excitation annihilation and the contributions of annihilation and quenching to the measured fluorescence kinetics can be distinguished even with ps-pulsed excitation. \par

\vspace{0.8cm}
\noindent \textbf{Declaration of competing interests} \par
The authors declare that no competing financial interests or personal relationships influenced the content of this work.\par
\vspace{0.8cm}
\noindent \textbf{Declaration of AI tool usage} \par
During the preparation of this work, the authors used ChatGPT-4o and Grammarly for light language editing and readability improvement. The authors reviewed and edited the output and take full responsibility for the content of the publication.\par
\vspace{0.8cm}
\noindent \textbf{Data availability}\par
Data will be made available by the authors on reasonable request.\par
\vspace{0.8cm}
\noindent \textbf{Acknowledgments}\par
The authors thank Joshua Botha for the development of the LabVIEW interface for performing raster scanning using the setup's piezoelectric nanopositioning stage.\par
\vspace{0.8cm}
\noindent \textbf{Funding Sources}\par
F.C. acknowledges support from the South African Quantum Technology Initiative (SA QuTI). B.v.H. acknowledges support from the University of Pretoria and the National Institute for Theoretical and Computational Sciences (NITheCS). T.P.J.K. acknowledges funding from the National Research Foundation (NRF), South Africa (grant nos. 137973 and 0403211945) and the Rental Pool Programme of the Council for Scientific and Industrial Research’s Photonics Centre, South Africa. \par

%% The Appendices part is started with the command \appendix;
%% appendix sections are then done as normal sections

\appendix
\section{Data Analysis}
\label{sec:data_analysis:appendix}
FCS analysis was performed by fitting a standard 3D diffusion model combined with a triplet blinking model to autocorrelation curves with the Levenberg-Marquardt algorithm, given as the autocorrelation function (ACF) over delay time, $\tau$, \cite{Schwille2000, PyCorrFit}:
\begin{equation}
    G_{1}(\tau) = G_{\infty} + \frac{1}{N} \frac{1}{1+\frac{\tau}{\tau_D}} \frac{1}{\sqrt{1+\frac{\tau}{k^2\tau_D}}} \left( \frac{1-T+Te^{\tau/\tau_t}}{1-T} \right),
    \label{eq:FCS_model_T+3D}
\end{equation}

\noindent where $G_{\infty}$ is the autocorrelation offset at large $\tau$ values, $N$ is the average number of diffusing particles detected, $\tau_D$ is the average diffusion time of the particles, $k$ is the eccentricity of the confocal volume, $T$ is the fraction of molecules in a non-emissive triplet state during the course of the measurement, and $\tau_t$ is the average triplet state lifetime. \par

Where more than one diffusion time was apparent, i.e., when the one-component model in Eq. \ref{eq:FCS_model_T+3D} yielded an insufficient curve fit, a two-component 3D diffusion model was used \cite{PyCorrFit}: \par

\vspace{0.5cm}
$G_{2}(\tau) = G_{\infty}$ $+$ $\frac{1}{N (f+\alpha (1-f))^2} \times$ 
\begin{equation}
      \left[ \frac{f}{1+\frac{\tau}{\tau_{D_1}}} \frac{1}{\sqrt{1+\frac{\tau}{k^2\tau_{D_1}}}} 
     + \alpha^2\frac{1-f}{1+\frac{\tau}{\tau_{D_2}}} \frac{1}{\sqrt{1+\frac{\tau}{k^2\tau_{D_2}}}}\right],
    \label{eq:FCS_model_3D+3D}
\end{equation}

\noindent where $F$ is the fraction of the species of particles with diffusion time $\tau_{D_1}$ and $\alpha$ is the ratio of molecular brightness of species 1 and 2. Here, $N = N_1 + N_2$ is the sum of the numbers of both species of particles detected. \par

The 3D diffusion coefficient depends on the size of the effective detection volume of the FCS setup, $V_{eff}$, which depends on the confocal volume given as \cite{Buschmann2007}
\begin{equation}
    V_{conf} = \left(\frac{\pi}{2}\right)^{3/2}{\omega_0}^2 z_0 = \left(\frac{1}{2}\right)^{3/2}V_{eff},
    \label{eq:confocal_volume}
\end{equation}

\noindent where $\omega_0$ is the lateral radius of the focus and $z_0$ is the axial radius. The eccentricity is thus $k=z_0/\omega_0$. See Supplementary Methods for further calibration details. \par 
The average number of particles detected in $V_{eff}$ over the course of a measurement, $\langle N \rangle$, is given by \cite{Buschmann2007}
\begin{equation}
    \langle N \rangle = \frac{N}{\chi^2},
    \label{eq:particle_no_corr}
\end{equation}

 \noindent where $\chi^2$ is the background correction factor given by
\begin{equation}
    \chi^2 = \left(1 + \frac{\langle b \rangle}{F}\right)^2,
    \label{eq:Background_corr}
\end{equation}

 \noindent where $\langle b \rangle$ is the average background count rate and $F= F^{det}-\langle b \rangle$ is the average count rate of fluorescence without the background. \par

Knowing the lateral radius $\omega_0$, the average diffusion coefficient of particles is
\begin{equation}
    D = \frac{{\omega_0}^2}{4\tau_D},
    \label{eq:diffusion_coefficient}
\end{equation}

 \noindent from which the average hydrodynamic radius of the particles is given by the Stokes--Eintein relation of Brownian motion,
\begin{equation}
    R_H = \frac{k_BT}{6\pi \eta D},
    \label{eq:hydrodynamic_radius}
\end{equation}

 \noindent where $k_B$ is Boltzmann's constant, $T$ is the temperature (295 K for our measurements), and $\eta$ is the dynamic viscosity of water at temperature $T$. \par

Measured IRFs were used to perform deconvolution fitting of the fluorescence decays with a two- or three-component exponential model: 
\begin{equation}
    F(t) = \sum_{i=1}^n A_i e^{-t/\tau_i},
    \label{eq:mult_comp_decay}
\end{equation}

 \noindent where $A_i$ is the normalized amplitude of a decay component, $\tau_i$, for $n=2$ or 3. \par
The average fluorescence lifetime was calculated as 
\begin{equation}
    \tau^{avg} = \sum_{i=1}^n A_i \tau_i.
    \label{eq:avg_lifetime}
\end{equation}

The relative effective fluorescence yield of LHCII in solution at a higher excitation intensity, $I_e$, compared to that at a lower excitation intensity, $I^{0}_e$, was calculated as follows:
\begin{equation}
    \frac{\Phi^{eff}}{\Phi^{eff}_0} = \frac{F}{F_0}\frac{I^{0}_e}{I_e},
    \label{eq:rel_yield}
\end{equation}

\noindent where $\Phi^{eff}_0$ is the fluorescence yield and $F_0$ the average fluorescence count rate of LHCII trimers when excited with $I^{0}_e$. \par

The relative fluorescence yield of LHCII at a constant excitation intensity during aggregation was calculated with $I^{0}_e=I_e$, in which case $F_0$ was the average fluorescence count rate of LHCII trimers (in 0.03\% (w/v) $\alpha$-DM) before Bio-Beads incubation started and $F$ the average count rate after some detergent removal time. \par

\section{Supplementary Materials}
\label{sec:supp_mat:appendix}
Supplementary methods and figures for this article can be found online at (URL).

%% If you have bibdatabase file and want bibtex to generate the
%% bibitems, please use
%%
%%\bibliographystyle{elsarticle-num} 
\bibliography{export_corrected.bib}

%% else use the following coding to input the bibitems directly in the
%% TeX file.

%%\printbibliography
%\begin{thebibliography}{00}

% %% \bibitem{label}
% %% Text of bibliographic item

% \bibitem{}

%\end{thebibliography}
\end{document}